\documentclass[useAMS,usenatbib]{mn2e}

\usepackage{amssymb}
\usepackage{graphicx}
\usepackage{url}

\def\aj{AJ}
\def\apj{ApJ}
\def\apjl{ApJ}
\def\apjs{ApJS}
\def\aap{A\&A}
\def\aaps{A\&AS}
\def\mnras{MNRAS}
\def\pasa{PASA}
\def\nat{Nature}

\newcommand\Msun{\,\rmn{M_\odot}}
\newcommand\MLsun{\,(\rmn{M/L})_\rmn{\odot}}

\newcommand\Mvir{M_\mathrm{vir}}
\newcommand\FeH{\lbrack \rmn{Fe}/\rmn{H} \rbrack}
\newcommand\ZH{\lbrack \rmn{Z}/\rmn{H} \rbrack}
\newcommand\alphaFe{\lbrack \alpha/\mathrm{Fe} \rbrack}

\title[Constraining UCD formation]{Constraining ultra-compact dwarf galaxy formation with galaxy clusters in the local universe}
\author[J. Pfeffer et al.]{J. Pfeffer$^{1,2}$\thanks{E-mail: j.l.pfeffer@ljmu.ac.uk}, M. Hilker$^{3}$, H. Baumgardt$^{1}$, B.~F. Griffen$^{4}$ \\
$^{1}$School of Mathematics and Physics, The University of Queensland, Brisbane, QLD 4072, Australia \\
$^{2}$Astrophysics Research Institute, Liverpool John Moores University, 146 Brownlow Hill, Liverpool L3 5RF, UK \\
$^{3}$European Southern Observatory (ESO), Karl-Schwarzschild-Strasse 2, 85748 Garching, Germany \\
$^{4}$Massachusetts Institute of Technology, Kavli Institute for Astrophysics and Space Research, 77 Massachusetts Avenue, Cambridge, \\ MA 02139, USA}

\begin{document}

\date{}

\pagerange{\pageref{firstpage}--\pageref{lastpage}} \pubyear{2016}

\maketitle

\label{firstpage}

\begin{abstract}
We compare the predictions of a semi-analytic model for ultra-compact dwarf galaxy (UCD) formation by tidal stripping to the observed properties of globular clusters (GCs) and UCDs in the Fornax and Virgo clusters. For Fornax we find the predicted number of stripped nuclei agrees very well with the excess number of GCs$+$UCDs above the GC luminosity function. GCs$+$UCDs with masses $>$10$^{7.3} \Msun$ are consistent with being entirely formed by tidal stripping. Stripped nuclei can also account for Virgo UCDs with masses $>$10$^{7.3} \Msun$ where numbers are complete by mass. For both Fornax and Virgo, the predicted velocity dispersions and radial distributions of stripped nuclei are consistent with that of UCDs within $\sim$50-100 kpc but disagree at larger distances where dispersions are too high and radial distributions too extended. Stripped nuclei are predicted to have radially biased anisotropies at all radii, agreeing with Virgo UCDs at clustercentric distances larger than 50 kpc. However, ongoing disruption is not included in our model which would cause orbits to become tangentially biased at small radii. We find the predicted metallicities and central black hole masses of stripped nuclei agree well with the metallicities and implied black hole masses of UCDs for masses $>$10$^{6.5} \Msun$. The predicted black hole masses also agree well with that of M60-UCD1, the first UCD with a confirmed central black hole. These results suggest that observed GC$+$UCD populations are a combination of genuine GCs and stripped nuclei, with the contribution of stripped nuclei increasing toward the high-mass end.
\end{abstract}


\section{INTRODUCTION}

\defcitealias{Pfeffer:2014}{P14}

Ultra-compact dwarf galaxies (UCDs) are a new type of galaxy discovered just over 15 years ago in spectroscopic surveys of the Fornax galaxy cluster with sizes ($R_\mathrm{e} \lesssim 100$ pc) and luminosities ($-14 \lesssim M_V \lesssim -12$) intermediate between globular clusters (GCs) and dwarf galaxies \citep{Hilker:1999a, Drinkwater:2000}. Follow-up high resolution spectroscopy has found that UCDs have similar internal velocity dispersions to dwarf elliptical nuclei \citep[$\sigma \sim$30 km s$^{-1}$,][]{Drinkwater:2003}. Initially labelled UCDs \citep{Phillipps:2001} or dwarf-globular transition objects \citep{Hasegan:2005} their formation mechanism is still under debate, however two main scenarios have been suggested: they may be the high-mass end of the GC mass function observed around galaxies with rich GC systems \citep*{Mieske:2002, Mieske:2012} or the nuclei of tidally stripped dwarf galaxies \citep*{Bekki:2001, Bekki:2003, Drinkwater:2003, Pfeffer:2013}.

Through dedicated surveys many new UCDs have been found \citep*{Mieske:2004a,Firth:2007,Firth:2008,Gregg:2009}, to the point where the confirmed objects now number in the hundreds in the Fornax cluster alone \citep{Mieske:2012}.
UCDs have since been discovered in many environments, including other galaxy clusters (Abell 1689: \citealp{Mieske:2004b}; Virgo: \citealp{Hasegan:2005,Jones:2006}; Centaurus: \citealp{Mieske:2007}; Coma: \citealp{Price:2009,Chiboucas:2010}; Hydra I: \citealp{Misgeld:2011}; Perseus: \citealp{Penny:2012}), galaxy groups (Dorado and NGC 1400: \citealp{Evstigneeva:2007a}; NGC 5128: \citealp{Rejkuba:2007}; HCG 22 and HCG 90: \citealp{DaRocha:2011}; NGC 3923: \citealp{Norris:2011}) and around isolated galaxies (NGC 7252: \citealp{Maraston:2004}; Sombrero: \citealp{Hau:2009}; NGC4546: \citealp{Norris:2011}).
As more UCDs are found there is growing evidence that no single formation mechanism is responsible for their formation, however most UCDs either formed as giant GCs or by tidal stripping of nucleated dwarf galaxies \citep{Hasegan:2005, Mieske:2006, Brodie:2011, Chilingarian:2011, DaRocha:2011, Norris:2011, Norris:2014, Pfeffer:2014}.
Although the formation mechanism of a few peculiar objects can be determined \citep[e.g.][]{Seth:2014, Norris:2015}, disentangling the individual formation mechanism of most UCDs is almost impossible due to the similar predictions of internal UCD properties from each formation scenario. Determining the origin of UCDs therefore requires detailed predictions of how much each possible formation mechanism contributes to UCD populations.

Tidal stripping of nucleated galaxies is a likely origin for many UCDs \citep[and a confirmed origin for two objects,][]{Seth:2014, Norris:2015} however their contribution to the total UCD population is uncertain (and partly compounded by the various definitions of UCDs).
Previous studies have shown that tidal stripping of nucleated galaxies can produce objects with similar properties to observed UCDs \citep{Bekki:2003, Pfeffer:2013}. A number of studies presented estimates for the number of UCDs formed due to tidal disruption \citep*{Bekki:2003, Goerdt:2008, Thomas:2008}. However as these estimates were based on UCD formation in clusters with static potentials they suffer from a number of problems.
Static models do not take into account UCD formation that may have occurred within subclusters that later fell into the main cluster. Since galaxy clusters are expected to undergo many mergers during their formation, galaxies in clusters may be on chaotic orbits providing a few close pericentre passages necessary for UCD formation but far from the cluster centre at other times \citep{Pfeffer:2013}\footnote{However this study neglected dark matter in the dwarf galaxies and used unrealistic orbits and therefore may overestimate the ability of tidal stripping to form UCDs.}. 
Galaxies orbiting in clusters may also have interactions with other satellite galaxies thereby making tidal disruption more likely.

In \citet[hereafter P14]{Pfeffer:2014} we presented the first model for UCD formation based on cosmological simulations of galaxy formation. Our model uses a semi-analytic galaxy formation model to select possible UCD progenitor galaxies and to determine when they become disrupted by tidal forces. Assuming that galaxies at high redshift have the same distribution of nucleus-to-galaxy mass and nucleation fraction as galaxies in the present day Universe we determined the numbers and masses of UCDs formed by tidal stripping. Some preliminary analysis was presented comparing the number of UCDs predicted with the observed number in the Fornax cluster, finding at most $\sim$10 per cent of UCDs have formed by tidal stripping.

In this paper we compare in detail the predictions of the \citetalias{Pfeffer:2014} model with the properties of UCDs from the Fornax and Virgo clusters. In particular we compare the predicted mass functions, radial and velocity distributions, metallicities and central black hole masses with the observed distributions.
Throughout the paper we refer to objects formed in the simulation by tidal stripping of nucleated galaxies as stripped nuclei since such objects may resemble both GCs and UCDs and because the observed UCD populations may be the result of more than one formation channel.

This paper is organized as follows. Section \ref{sec:SAM} describes the criteria for selecting analogue galaxy clusters of the Fornax and Virgo clusters and briefly summarises the method of \citetalias{Pfeffer:2014} for identifying stripped nuclei in cosmological simulations. Section \ref{sec:observations} describes the compilation of observational data of GCs, UCDs and dwarf galaxies. Section \ref{sec:results} presents the results comparing the simulation and observational data. In Section \ref{sec:discussion} and \ref{sec:summary} we discuss the implications of our work for UCD formation scenarios and summarize our results.


\section{Semi-analytic modelling} \label{sec:SAM}

\defcitealias{Guo:2011}{G11}

Here we summarize the stripped nucleus formation model of \citetalias{Pfeffer:2014} and detail our selection criteria for comparing against observed galaxy clusters.

The model makes use of the semi-analytic galaxy formation model (SAM) of \citet[hereafter G11]{Guo:2011} which was applied to the subhalo merger trees of the Millennium-II simulation \citep[hereafter MS-II]{Boylan-Kolchin:2009}. The MS-II is a cosmological dark-matter only simulation which has a box size of 137 Mpc and a particle mass of $9.42\times 10^6 \Msun$. The \citetalias{Guo:2011} SAM is constrained by low-redshift galaxy abundance and clustering in the Sloan Digital Sky Survey and is tuned to reproduce the $z=0$ mass distribution of galaxies down to stellar masses of $10^{7.5} \Msun$. 
In the \citetalias{Guo:2011} SAM satellite galaxies (i.e. those currently or previously at the centre of non-dominant haloes orbiting within a more massive halo) may have either resolved or unresolved dark matter (DM) haloes under the assumption that the stellar component of a satellite galaxy is harder to disrupt than its halo. For the stellar component of a satellite galaxy to be disrupted its DM halo must first be entirely dissolved (i.e. become unresolved). Since tidal stripping is not taken into account in the model, satellite galaxies do not lose stellar mass until they are completely disrupted.
For all data associated with MS-II and the \citetalias{Guo:2011} SAM we assume a cosmology consistent with the \textit{Wilkinson Microwave Anisotropy Probe} 1-year data (WMAP1) results \citep{Spergel:2003} and assume $h=0.73$ for all masses and distances\footnote{In \citetalias{Pfeffer:2014} we tested semi-analytic models for both a WMAP1 \citep{Guo:2011} and WMAP7 cosmology \citep{Guo:2013}, finding no significant difference between the predictions. Therefore modelling with an updated cosmology would not change our results.}. The data associated with the MS-II and \citetalias{Guo:2011} SAM are publicly provided by the Virgo-Millennium Database \citep{Lemson:2006}\footnote{\url{http://www.mpa-garching.mpg.de/millennium}}.

In \citetalias{Pfeffer:2014} a sample of galaxy clusters was chosen such that $\Mvir > 10^{13} \Msun/h$. Here we limit the cluster sample according to the mass of the cluster we are comparing with. For the Fornax cluster we choose all clusters within the range $\Mvir = 7\pm2 \times 10^{13} \Msun$ \citep{Drinkwater:2001}, giving 37 clusters for comparison. For the Virgo cluster we choose all clusters within the range $\Mvir = (1.4$--$7) \times 10^{14} \Msun$ \citep{McLaughlin:1999, Tonry:2000, Urban:2011}, giving 13 clusters for comparison.

After the SAM clusters are chosen, stripped nuclei are identified in the simulations in the following way:

\begin{enumerate}

\item The galaxy merger trees of all galaxies in the SAM clusters at $z=0$ are searched for possible stripped nucleus progenitors (hereafter referred to as candidate galaxies and the DM halo of the galaxies as candidate haloes). A galaxy is defined as a possible progenitor when the stellar mass exceeds $10^{7.5} \Msun$ (i.e. all progenitors of the candidate galaxy have a stellar mass less than this limit). This lower mass cut is chosen based on the lower mass limit observed for nucleated galaxies \citepalias[see figure 1 of][]{Pfeffer:2014}.

\item To form a stripped nucleus the candidate galaxy must undergo a merger with a more massive galaxy and the stellar component of the galaxy be completely disrupted according to the criteria in the SAM (equation 30 of \citetalias{Guo:2011}). It is assumed that during this process the nuclear cluster of the galaxy is not disrupted since it is much more compact than the galaxy \citep{Bekki:2003, Pfeffer:2013}.

\item The galaxy merger must be a minor merger\footnote{It is important to note that in the minor merger process we expect the candidate galaxy was disrupted by the tidal field of the host galaxy. In this case the nuclear cluster of the galaxy is not disrupted and forms a UCD \citep{Bekki:2003, Pfeffer:2013}. In the case of a major merger the nuclei of the two galaxies merge and a UCD is not produced.}, where a minor merger is defined as those with `dynamical' mass ratios smaller than 1:3. The dynamical mass $M_\mathrm{dyn}$ is defined as $M_\mathrm{dyn} = M_* + M_\mathrm{gas} + M_\mathrm{DM}(<r_\mathrm{s})$ where $M_*$ is the stellar mass of the galaxy, $M_\mathrm{gas}$ is the cold gas mass and $M_\mathrm{DM}(<r_\mathrm{s})$ is the mass of the DM halo within the NFW scale radius.

\item The candidate galaxy merged\footnote{Galaxy mergers in the model are assumed to happen on a time scale shorter than the time between simulation snapshots (a maximum of 377 Myr) which may not be true in reality.} (i.e. the stellar component of the galaxy was disrupted according to the SAM) at least 2 Gyr ago so there is enough time to form a UCD \citep{Pfeffer:2013}.

\item The dynamical friction time of the stripped nucleus $t_\mathrm{friction}$ \citep[determined using equation 7-26 from][]{Binney:1987} is calculated at the snapshot before the candidate galaxy is disrupted. After a time $t_\mathrm{friction}$ the stripped nucleus is assumed to merge with the centre of the galaxy it is orbiting and therefore be completely disrupted. This differs from the treatment of dynamical friction in \citetalias{Pfeffer:2014} where the dynamical friction time was calculated for the halo the stripped nucleus is associated with at $z=0$. We find the predictions of the model are unaffected by this change.

\end{enumerate}

After the candidate galaxy is disrupted the stripped nucleus takes the position and velocity of the most bound particle of the candidate halo (similar to galaxies without resolved haloes in the \citetalias{Guo:2011} SAM). 
We also take into account orbital decay of the stripped nuclei due to dynamical friction by modifying the distance of the stripped nuclei from their host galaxies by a factor $(1-\Delta t/t_\mathrm{friction})$, where $\Delta t$ is the time since the progenitor galaxy was disrupted (this is identical to the treatment of satellite galaxies in the \citetalias{Guo:2011} SAM). The velocity of the stripped nucleus relative to the host galaxy was kept the same as the velocity of the most bound particle relative to the host halo (to account for galaxies undergoing orbital decay in the SAM).

The two main assumptions for assigning physical properties to the stripped nuclei are the following:
\begin{enumerate}

\item The stripped nucleus is assigned a mass randomly chosen from a log-normal mass function for the nucleus-to-galaxy mass ratio $M_\mathrm{nuc}/M_\mathrm{*,gal}$ (where the mass of the candidate galaxy is taken immediately before disruption).
Based on figure 14 from \citet{Cote:2006}, we choose a mean of 0.3 per cent and a log-normal standard deviation of 0.5 dex.
Tidal stripping of the stripped nuclei is not taken into account.
Recently, \citet{Georgiev:2016} found $M_\mathrm{nuc}/M_\mathrm{*,gal} \simeq 0.1$ per cent in both early- and late-type galaxies, lower than the value we assumed. Therefore when comparing mass functions in Section \ref{sec:massFunction} we also consider this value for the nucleus-to-galaxy mass ratio.

\item The fraction of galaxies that are nucleated is taken from observations of galaxies in the Virgo and Fornax clusters \citepalias[see figure 1 of][]{Pfeffer:2014}. 
For galaxies between stellar masses of $3.0 \times 10^7$ and $4.7 \times 10^8 \Msun$ we choose a nucleation fraction that varies linearly (in log-space) between 0 and 80 per cent. Between stellar masses of $4.7 \times 10^8$ and $10^{11} \Msun$ we take an average nucleation fraction of 80 per cent. For galaxies above stellar masses of $10^{11} \Msun$ we assume nuclear clusters are destroyed by supermassive black holes \citep[e.g.][]{Graham:2009}.
Where possible, to improve our statistics we work with fractions of stripped nuclei instead of randomly choosing galaxies to satisfy the nucleated fraction.

\end{enumerate}


\defcitealias{Maraston:2005}{M05}
\defcitealias{Bruzual:2003}{BC03}

\section{Observational data} \label{sec:observations}

For all data associated with the Fornax cluster we use the distance modulus $m-M = 31.39$ mag \citep{Freedman:2001} corresponding to a distance of 19 Mpc and a spatial scale of 92 pc arcsec$^{-1}$.
For all data associated with the Virgo cluster we use the distance modulus $m-M = 31.09$ mag \citep{Mei:2007} corresponding to a distance of 16.5 Mpc and a spatial scale of 80 pc arcsec$^{-1}$.

\subsection{Fornax cluster} \label{sec:FornaxData}

\subsubsection{GCs and UCDs}

Since UCDs were discovered, the GC and UCD population of the Fornax cluster has been the subject of many studies. To determine the number of spectroscopically confirmed GCs$+$UCDs in Fornax we compile the number and luminosities of objects from \citet{Hilker:1999a}, \citet{Drinkwater:2000}, \citet{Mieske:2002, Mieske:2004a}, \citet{Bergond:2007}, \citet{Hilker:2007}, \citet{Firth:2007}, \citet{Mieske:2008}, \citet{Gregg:2009}, \citet{Schuberth:2010}, \citet{Chilingarian:2011} and Puzia \& Hilker (priv. comm.). The total number of confirmed GCs$+$UCDs in the Fornax cluster is 935.
Due to the 2dF Fornax surveys of \citet{Drinkwater:2000} and \citet{Gregg:2009}, the number of GCs$+$UCDs with luminosities brighter than $M_V < -11.5$ mag are more than 95 per cent complete within a clustercentric radius of 0.9 degrees, or $\sim$300 kpc.
As most studies concentrated on the inner 15 acrmin of the cluster, or $<$83 kpc, the number of GCs$+$UCDs with luminosities $M_V < -10.25$ is approximately complete within 50 kpc and the completeness drops to below 70 per cent beyond this radius \citep[see discussion in][]{Mieske:2012}.

To compare the model predictions with the observed population it is necessary to derive masses for the GCs and UCDs from population synthesis models. 
GCs are generally very old \citep[$\sim 13$ Gyr,][]{Marin-Franch:2009} and are consistent with having a Kroupa/Chabrier stellar initial mass function \citep[IMF; e.g.][]{McLaughlin:2005, Paust:2010}. Note that GCs actually show a spread in MFs due to ongoing dissolution, however at the high mass end for GCs which we are interested in this is not important \citep{Baumgardt:2003}. 
UCDs on the other hand are on average slightly younger than GCs \citep[$\sim 11$ Gyr,][]{Francis:2012}. 
Many UCDs have dynamical mass-to-light ($M/L$) ratios above what can be explained by stellar population models \citep{Hasegan:2005, Dabringhausen:2008, Mieske:2013}. One suggestion is that the elevated $M/L$ ratios are due to a top-heavy IMF \citep*[although see also \citealt{Phillipps:2013}]{Dabringhausen:2009, Dabringhausen:2012}. Therefore an IMF leading to higher masses should also be considered.
To determine the effect of the assumed stellar population on our results we consider a number of different IMFs and simple stellar population (SSP) models. 
We use the \citet[M05]{Maraston:2005} SSP model with \citet{Kroupa:2001} and \citet{Salpeter:1955} IMFs and the \citet[BC03]{Bruzual:2003} SSP model with a \citet{Chabrier:2003} IMF. Each SSP model is generated for two different ages, 11 and 13 Gyr. 
For the \citetalias{Maraston:2005} models we tested both blue (bHB) and red horizontal branch (rHB) models but find no difference between the results and therefore only consider the rHB models hereafter.
Masses for the GCs$+$UCDs were then calculated from the $(V-I)$-$M/L_V$ relation fitted to the different SSP models \citep[see also][]{MisgeldHilker:2011}.
Given the observed properties of GCs and UCDs the Kroupa/Chabrier SSP models are the most reasonable. The Salpeter IMF (which is a bottom-heavy IMF compared to Kroupa/Chabrier) represents the case of elevated $M/L$ ratios. In terms of mass, the completeness of $M_V < -11.5$ mag within 300 kpc corresponds to $M \gtrsim 10^{6.9} \Msun$ for the Kroupa/Chabrier SSP models. 

To estimate the completeness of the Fornax sample within 83 kpc in terms of mass we compare the number of spectroscopically confirmed GCs$+$UCDs with the GC luminosity function (GCLF, derived from photometry, where the errors come from uncertainties in background corrections) within 83 kpc \citep[$6,450\pm700$ GCs,][]{Dirsch:2003} and 300 kpc \citep[$11,100\pm2,400$ GCs,][derived from the data of \citealt{Bassino:2006}]{Gregg:2009}. It is assumed that the GCLF has a Gaussian shape with a width $\sigma=1.3$ mag and peak $M_V = -7.6$ mag \citep{Dirsch:2003}. We convert the GCLF to a mass function for a given SSP model/IMF assuming a constant $M/L$ ratio given by median of the GC$+$UCD sample (to be dominated by objects with masses $\sim 10^6 \Msun$) using 13 Gyr ages. 
In Fig. \ref{plt:fornaxMF} we compare the number of spectroscopically confirmed GCs$+$UCDs with the GCLF for each SSP model. Comparing the results of all SSP models for the GC+UCD sample with the GCLF we estimate a completeness limit of $\sim 10^{6.7} \Msun$ within 83 kpc.

\subsubsection{Dwarf galaxies}

We take the sample of Fornax dwarf galaxies from the Fornax Cluster Catalogue \citep[FCC,][]{Ferguson:1989} with updated radial velocities from \citet{Thomas:2008}. 
We exclude one galaxy (FCC2) as a background galaxy due to its high line-of-sight velocity ($cz = 4540$ km s$^{-1}$) compared to the cluster \citep[$cz \sim 1500$ km s$^{-1}$,][]{Gregg:2009}.
The catalogue is complete to $B_\mathrm{T} \sim 18$ mag, or $M_B \sim -13.4$ mag, and therefore we take this as the lower luminosity limit for the sample. Assuming a stellar mass-to-light ratio $M/L_B = 3 \MLsun$ \citep[typical for dwarf ellipticals,][]{Chilingarian:2009} this corresponds to a stellar mass $M_* = 10^8 \Msun$.
A galaxy is considered to be a dwarf galaxy if it has a luminosity $M_B > -19.6$, corresponding to $M < 10^{10.5} \Msun$ for $M/L_B = 3 \MLsun$.
Therefore when comparing against dwarf galaxies in the FCC, we choose dwarf galaxies in the simulations that have stellar masses $10^8 < M_*/\Msun < 10^{10.5}$.

\subsection{Virgo cluster}

\subsubsection{GCs and UCDs}

The recent Next Generation Virgo cluster Survey \citep[NGVS,][]{Ferrarese:2012}, together with spectroscopic follow-up \citep{Zhang:2015}, allows for a systematic comparison with UCDs in the Virgo cluster. Virgo cluster UCDs, specifically UCDs around the central galaxy M87, were therefore taken from \citet{Zhang:2015}. 
Objects selected from the NGVS have luminosities $M_g \leq -9.6$ mag and effective radii $R_e \geq 11$ pc, while previously confirmed objects selected from HST imaging \citep{Brodie:2011} have effective radii $R_e > 9.5$ pc. All UCDs selected have an upper size limit $R_e \lesssim 100$ pc. The total number of UCDs in the \citet{Zhang:2015} catalogue is 97. The number of UCDs is expected to be nearly 100 per cent complete within 288 kpc for $g < 20.5$ mag.
However, in the \citet{Zhang:2015} study UCDs are defined by effective radius ($10 \lesssim R_e/\mathrm{pc} \lesssim 100$) while in our model we can currently only compare objects by mass. With this shortcoming in mind, in this work we give basic comparisons between Virgo UCDs and our model. More detailed analysis therefore requires observations that are complete above a given mass or a model which can predict stripped nuclei sizes. 

To derive masses for the \citet{Zhang:2015} sample of UCDs we follow the method of \citet{MisgeldHilker:2011}. The apparent $g_\mathrm{AB}$-band magnitudes were transformed into absolute $V$-band magnitudes using the relation $V = g_\mathrm{AB} + 0.026 - 0.307 \times (g − z)_\mathrm{AB}$ given in \citet{Peng:2006} and a Virgo distance modulus of 31.09 mag \citep{Mei:2007}. Mass-to-light ratios $M/L_V$ were calculated assuming a $M/L_V$-$(g-z)_\mathrm{AB}$ relation according to \citet{Maraston:2005} single stellar population models (Kroupa IMF, red horizontal branch) for ages of 11 and 13 Gyr \citep{MisgeldHilker:2011}.
For objects H30772 and S887 we take $z$-band magnitudes from \citet{Jordan:2009}.
For the luminosity completeness limit $g < 20.5$ mag, this corresponds to a completeness in mass of $M \gtrsim 10^{6.6} \Msun$.

To compare velocity dispersions of stripped nuclei with GCs we use the sample of GCs compiled by \citet[excluding any UCDs defined in \citealt{Zhang:2015}]{Strader:2011}.
For comparing radial number profiles of GCs around M87 we use the S\'ersic profiles fitted by \citet{Zhang:2015} to the surface number density profiles for blue {[}$0.55 < (g^{\prime} - i^{\prime})_0 < 0.80${]} and red {[}$0.80 < (g^{\prime} - i^{\prime})_0 < 1.15${]} GCs ($20 < g^{\prime}_0 < 24$) determined by \citet{Durrell:2014} using NGVS photometry. The profiles for the blue and red GCs were fitted within 240 and 120 kpc from M87, respectively.

\subsubsection{Dwarf galaxies}

For comparing the radial distribution of Virgo cluster galaxies we take the sample of dwarf galaxies from the Virgo Cluster Catalogue \citep[VCC,][excluding galaxies classified as background galaxies]{Binggeli:1985}. The catalogue is complete to $B_\mathrm{T} \sim 18$ mag, or $M_B \sim -13.1$ mag. As for the Fornax cluster, we assume a typical mass-to-light ratio of $M/L_B = 3 \MLsun$ and use a  stellar mass cut when comparing with simulations of $10^8 < M_*/\Msun < 10^{10.5}$.

For comparing velocities of Virgo cluster galaxies we take the sample of dwarf galaxies from the Extended Virgo Cluster Catalogue \citep[EVCC,][]{Kim:2014}.
Masses for the dwarf galaxies were again estimated following the method of \citet{MisgeldHilker:2011}. The apparent $g_\mathrm{AB}$-band magnitudes were transformed into absolute $V$-band magnitudes using the relation given in \citet{Peng:2006}. Mass-to-light ratios $M/L_V$ were calculated assuming a $M/L_V$-$(g-z)_\mathrm{AB}$ relation according to \citet{Maraston:2005} single stellar population models (Kroupa IMF, blue horizontal branch) for ages of 11 Gyr.
For comparison with the simulations we use a mass cut of $10^8 < M_*/\Msun < 10^{10.5}$.

\subsection{GC and UCD metallicities} \label{sec:obsMetallicities}

We compile a list of GC and UCD metallicities for Fornax and Virgo cluster GCs and UCDs from \citet{Mieske:2008}, \citet{Chilingarian:2011} and \citet{Francis:2012}. Where possible we use the dynamical masses measured for objects in this sample.
For common objects between the \citet{Mieske:2008} and \citet{Chilingarian:2011} samples we take data from \citeauthor{Chilingarian:2011}. For common objects between the \citet{Mieske:2008} and \citet{Francis:2012} samples we take data from \citeauthor{Francis:2012}. 
For the \citeauthor{Chilingarian:2011} sample we use the dynamical mass-to-light ratios to calculate the UCD masses where possible.
For the \citeauthor{Francis:2012} sample we take dynamical masses from \citet{Mieske:2013} for objects UCD1, UCD5, VUCD3 and VUCD5. For the other objects masses were calculated using the $g$ and $r$ colours and the stellar mass-to-light relations of \citet{Bell:2003}. For NGC1407-GC1 the $g$ and $r$ colours were taken from \citet{Romanowsky:2009}.
For NGC1407 we assume a distance modulus of 31.99 mag \citep*{Jerjen:2004}.


\section{RESULTS} \label{sec:results}

\subsection{Mass function} \label{sec:massFunction}

\begin{figure*}
  \centering
  \includegraphics[width=0.99\textwidth]{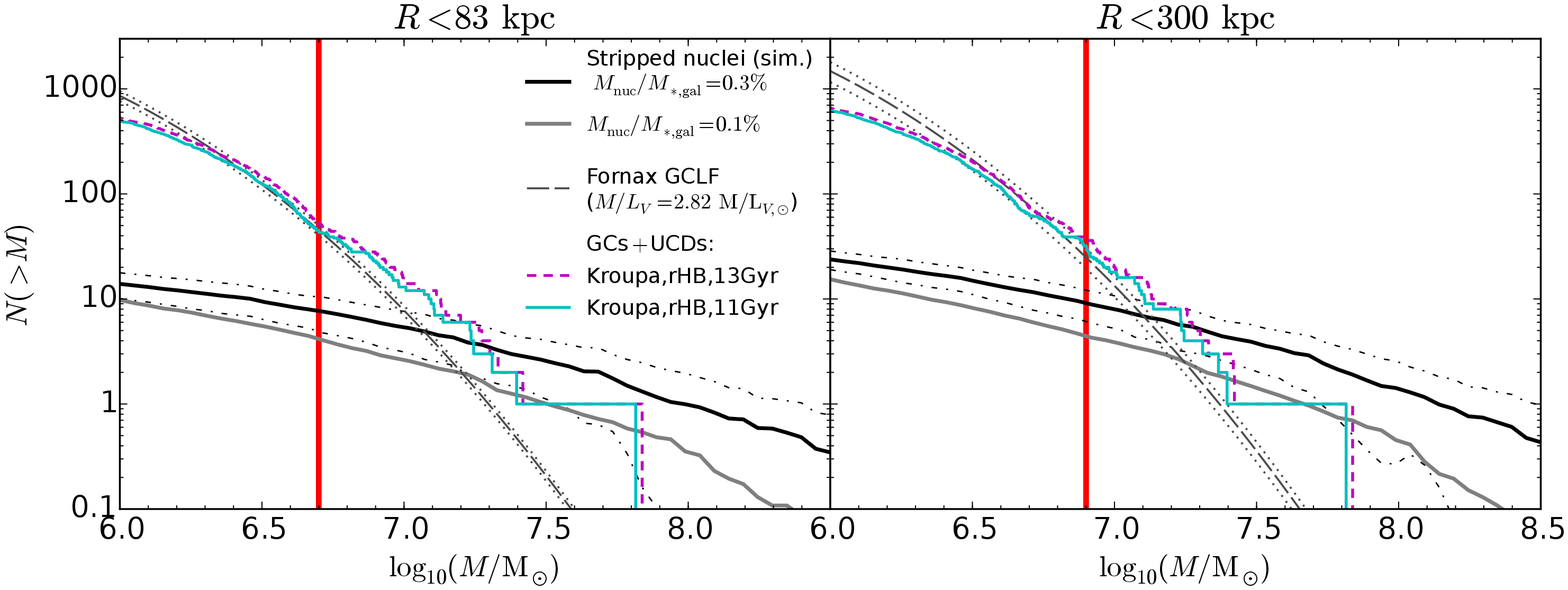}
  \includegraphics[width=0.99\textwidth]{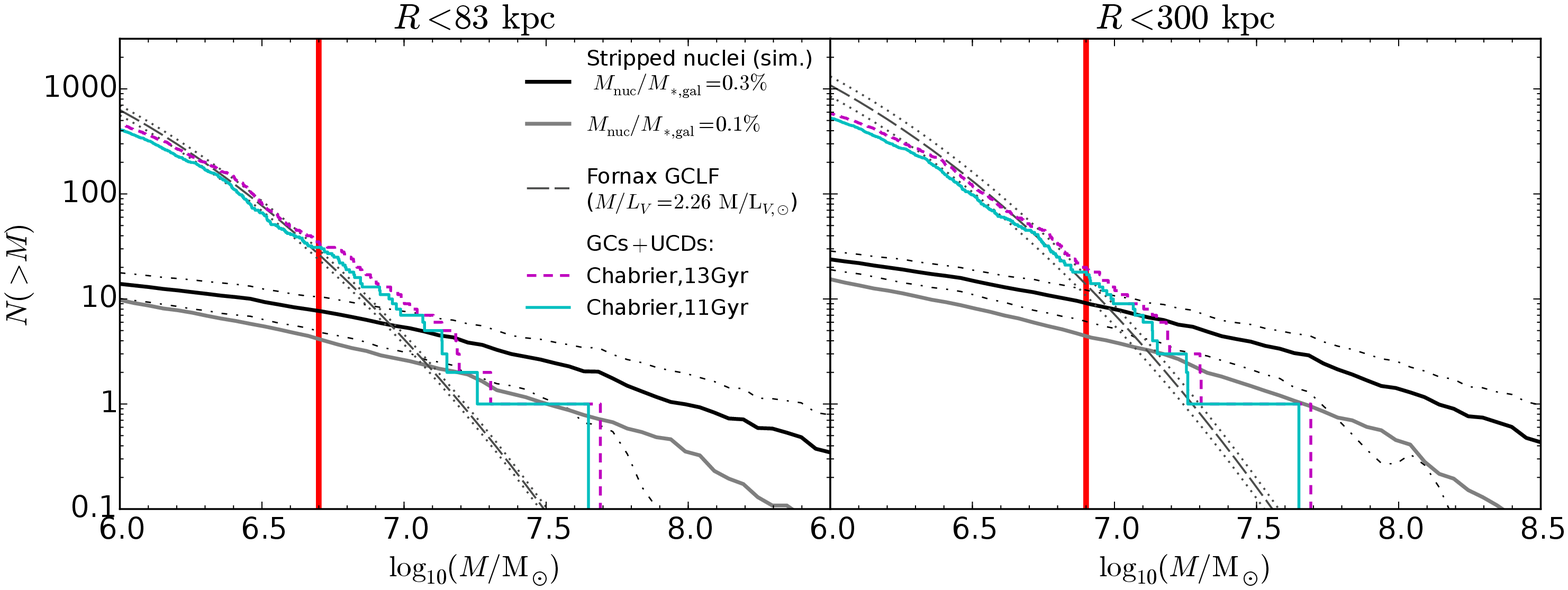}
  \includegraphics[width=0.99\textwidth]{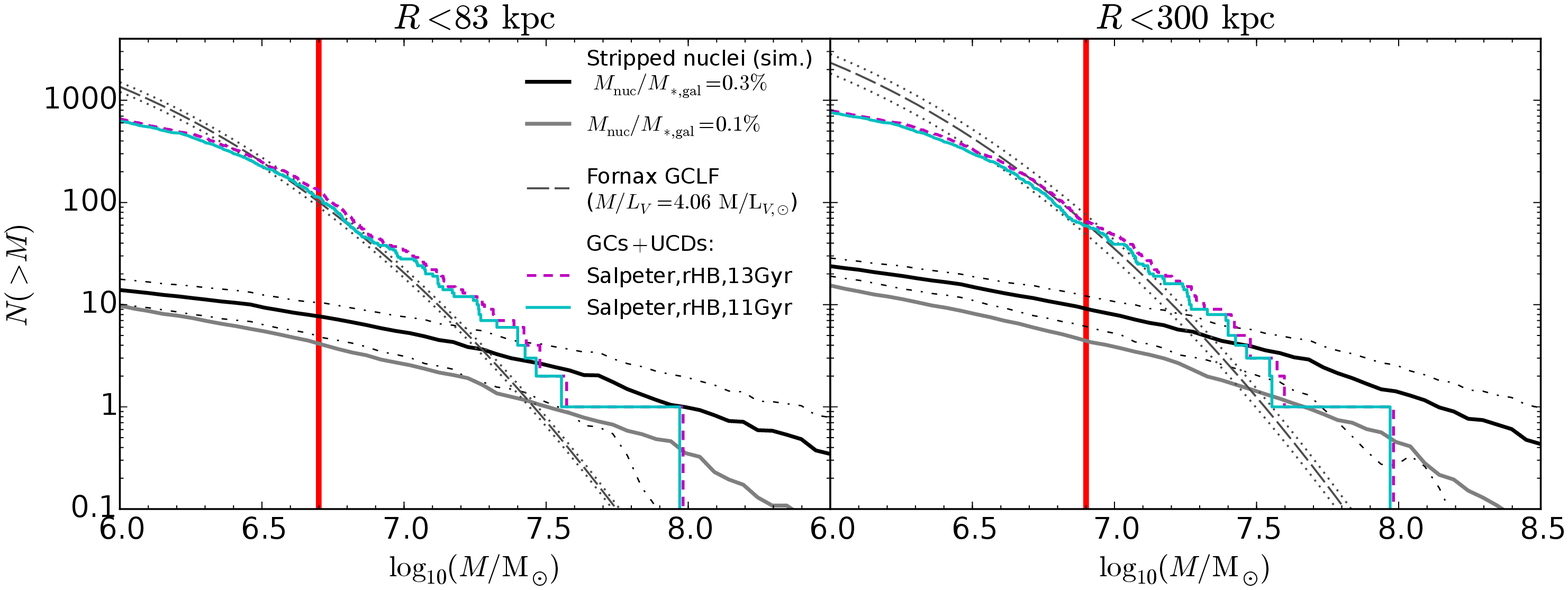}
  \caption{Cumulative mass function of simulated stripped nuclei in Fornax-like clusters assuming $M_\mathrm{nuc}/M_\mathrm{*,gal}$ is 0.3 per cent (black line, with standard deviation between clusters shown by black dash-dotted lines) and 0.1 per cent (grey line) compared with the number of spectroscopically confirmed Fornax GCs$+$UCDs (solid line, from Section \ref{sec:FornaxData}) and the cumulative Fornax GC luminosity function (GCLF, long-dashed line with the error shown by dotted lines) within 83 kpc \citep[$6,450\pm700$ GCs,][]{Dirsch:2003} and 300 kpc of NGC 1399 ($11,100\pm2,400$ GCs, \citealt{Gregg:2009}, derived from the data of \citealt{Bassino:2006}). The top panel shows the results for the \citet{Maraston:2005} SSP model with a Kroupa IMF, the middle panel for the \citet{Bruzual:2003} SSP model with a Chabrier IMF and the bottom panel for the \citet{Maraston:2005} SSP model with a Salpeter IMF. In each panel we show the results for the Fornax GC$+$UCD sample with SSP ages of 13 (dashed magenta lines) and 11 Gyr (solid cyan lines). The luminosity functions were converted to mass functions in each panel using the median $M/L$ ratio for the 13 Gyr SSP model (the $M/L$ ratio is given in the legend of each panel). The vertical red lines show the estimated completeness of the confirmed GCs$+$UCDs.}
  \label{plt:fornaxMF}
\end{figure*}

In Fig. \ref{plt:fornaxMF} we compare the cumulative mass function of simulated stripped nuclei in Fornax-like clusters with the number of confirmed Fornax GCs$+$UCDs compiled in Section \ref{sec:FornaxData} and the GCLF within 83 kpc \citep{Dirsch:2003} and 300 kpc (\citealt{Gregg:2009}, derived from the data of \citealt{Bassino:2006}) for each SSP model. The luminosity functions were converted to mass functions using a mass-to-light ratio given by the median of the Fornax GC$+$UCD sample for a given SSP model/IMF with an age of 13 Gyr. The luminosity function is approximated as a Gaussian with a width $\sigma=1.3$ mag and peak magnitude of $M_V=-7.6$ mag \citep{Dirsch:2003} with a total of $6,450\pm700$ GCs within 83 kpc \citep{Dirsch:2003} and $11,100\pm2,400$ GCs within 300 kpc \citep{Gregg:2009}. 
The figure suggests an excess of $\sim 10$ GCs$+$UCDs with masses $\gtrsim 10^7 \Msun$ compared to the GC mass function. As noted by \citet{Gregg:2009}, given the low number of objects at the high mass/luminosity end, the number of UCDs is consistent with being the high-mass tail of the GC mass function. 
The most massive Fornax UCD (UCD3) is also very extended \citep[$R_\mathrm{eff} = 90$ pc,][]{Hilker:2007} while all other Fornax UCDs have sizes less than 30 pc \citep[e.g.][]{MisgeldHilker:2011}. This may indicate UCD3 is not part of the GC mass function even if its mass is consistent with being part of it \citep[although GCs$+$UCDs formed by hierarchical merging of star clusters in cluster complexes may also reach such sizes,][]{Bruens:2012}.
However, unlike M60-UCD1 (the most massive UCD in the Virgo cluster) which has a supermassive black hole (SMBH) containing 15 per cent of its mass and which was most likely formed by tidal stripping \citep{Seth:2014}, UCD3 does not seem to have a SMBH \citep[at least not one containing more than 5 per cent of its mass,][]{Frank:2011}.

As already noted in \citetalias{Pfeffer:2014}, for masses more than $10^6 \Msun$ the number of stripped nuclei predicted by the model is a factor $>20$ smaller than the number of observed GCs$+$UCDs ($\sim 500$, while the number inferred from the GCLF is $\sim 1000$).
For masses between $10^6$ and $10^7 \Msun$ and $M_\mathrm{nuc}/M_\mathrm{*,gal} = 0.3$ per cent we predict that stripped nuclei account for $\sim2.5$ per cent of GCs$+$UCDs.
For masses larger than $10^7 \Msun$ we predict stripped nuclei account for $\sim40$ per cent of GCs$+$UCDs and become the dominant formation mechanism for masses larger than $\sim 10^{7.1}\Msun$ for a Kroupa/Chabrier IMF.
When assuming $M_\mathrm{nuc}/M_\mathrm{*,gal} = 0.1$ per cent the predicted number of stripped nuclei decreases by a factor $\sim$2.
As the number of dwarf galaxies at the centre of galaxy clusters is over-predicted by 50 per cent in the \citetalias{Guo:2011} SAM \citep[see also Fig. \ref{plt:fornaxRadialDist}]{Weinmann:2011}, the number of predicted stripped nuclei might increase by a factor of two if more efficient tidal processes were implemented in the simulations\footnote{Although the over-abundance of dwarf galaxies might also be related to the too efficient star formation in low-mass galaxies at early times or the over-clustering of galaxies at scales below 1 Mpc \citep{Guo:2011}. See section 5.2 in \citetalias{Pfeffer:2014} for a discussion about how this affects our predictions.}. However this is still too low to account for all observed UCDs.

\begin{figure*}
  \centering
  \includegraphics[width=0.99\textwidth]{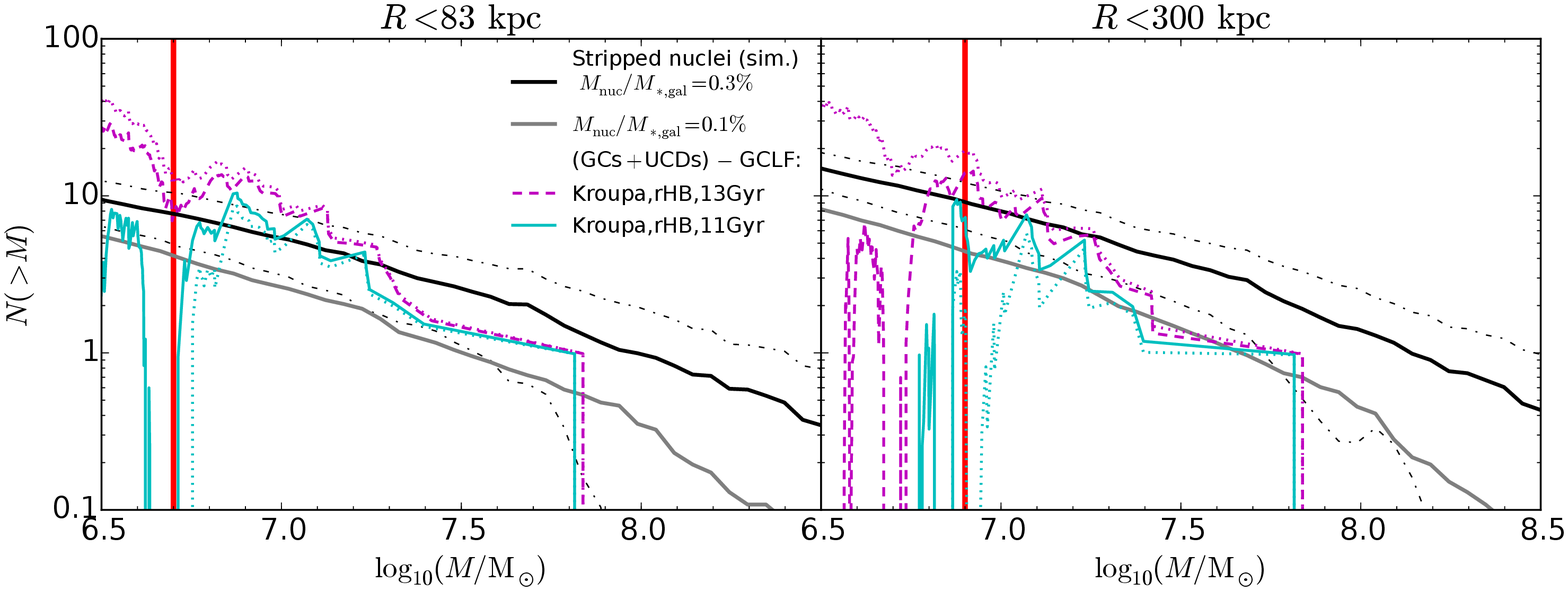}
  \includegraphics[width=0.99\textwidth]{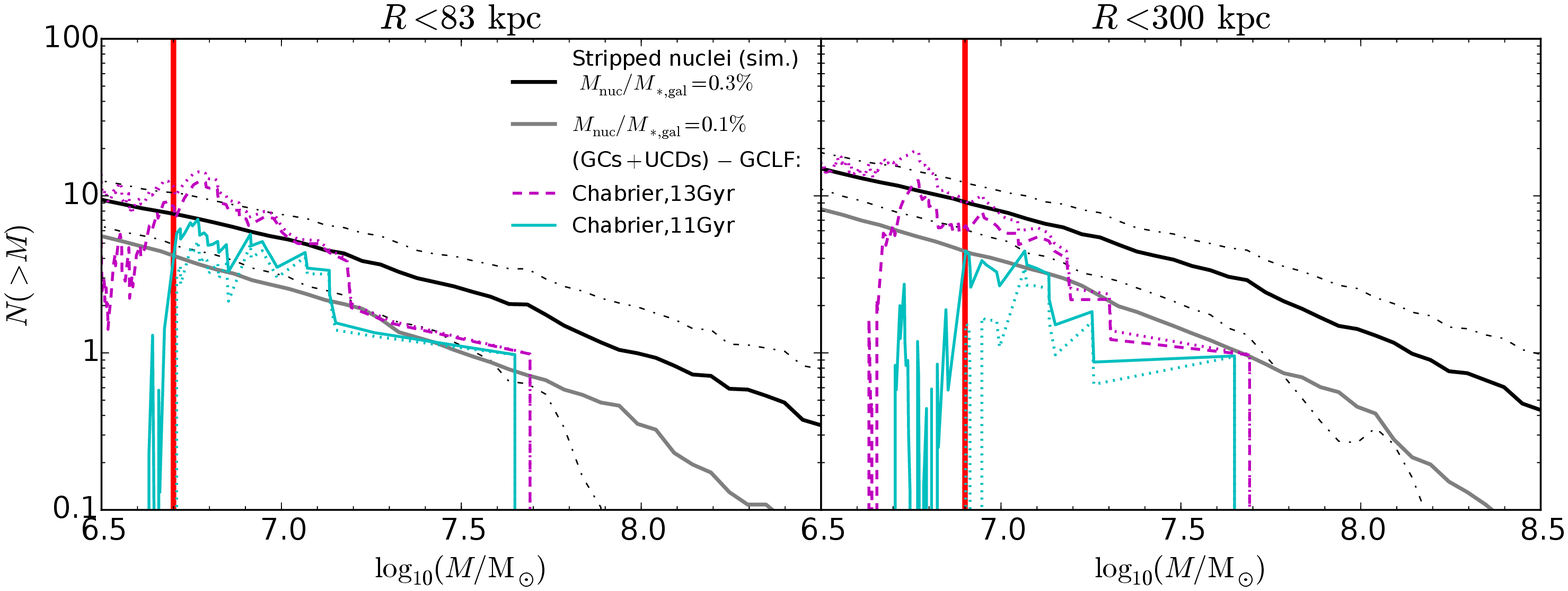}
  \includegraphics[width=0.99\textwidth]{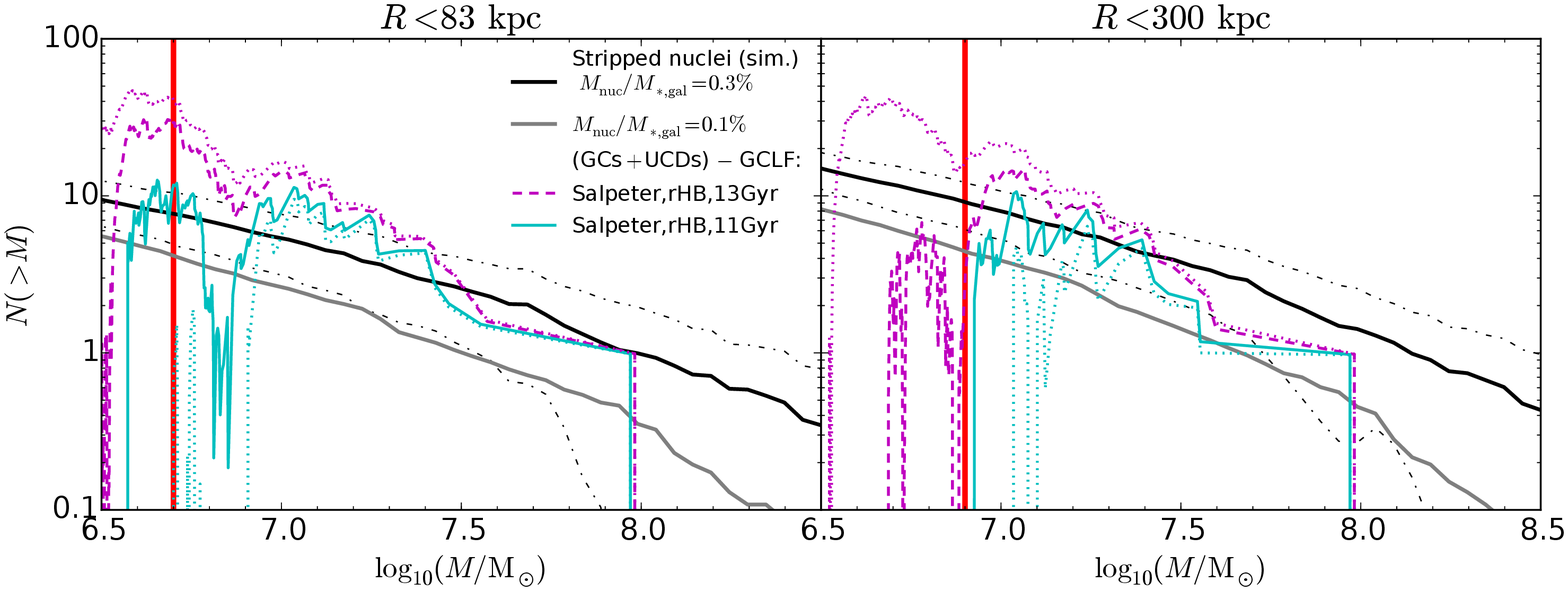}
  \caption{Cumulative mass function of simulated stripped nuclei in Fornax-like clusters within a projected distance of 83 (left panel) and 300 kpc (right panel) from the central galaxy assuming $M_\mathrm{nuc}/M_\mathrm{*,gal}$ is 0.3 per cent (black line, with standard deviation between clusters shown by black dash-dotted lines) and 0.1 per cent (grey line) compared with the difference of the Fornax GC$+$UCD sample and the GCLF {[}(GCs+UCDs)$-$GCLF{]} from Fig. \ref{plt:fornaxMF}. In each panel the results for SSP ages of 13 and 11 Gyr are shown by dashed magenta lines and solid cyan lines, respectively. 
The dotted lines show the upper limit for the 13 Gyr SSP (magenta) and lower limit for the 11 Gyr SSP (cyan) due to the error in the GCLF (the lower limit for the 13 Gyr SSP and upper limit for the 11 Gyr SSP are omitted for clarity).
Vertical red lines show the estimated completeness of the confirmed GCs$+$UCDs.}
  \label{plt:fornaxCMF}
\end{figure*}

In Fig. \ref{plt:fornaxCMF} we show the cumulative mass function for simulated stripped nuclei in Fornax-sized galaxy clusters compared with the difference of the number of spectroscopically confirmed GCs$+$UCDs with the GCLF for each SSP model: i.e. the excess number of objects which cannot be accounted for by the GCLF. The number of excess GCs$+$UCDs above the GCLF is slightly higher within 83 kpc than within 300 kpc which may indicate the number of objects beyond 83 kpc is not complete.
Overall, given the uncertainties in our model, the SSP models and the GCLF, the agreement between our model and the excess number of GCs$+$UCDs in the Fornax cluster is very good.
Within 83 kpc the predicted number of stripped nuclei agree well with the excess number of GCs$+$UCDs for all SSP models. Within 300 kpc the predicted number of stripped nuclei agree well with the number of GCs$+$UCDs above the GCLF for the Salpeter IMF. For the Kroupa and Chabrier IMFs the number of stripped nuclei predicted is too large above masses of $\sim 10^{7.3} \Msun$.
Above masses of $\sim 10^8 \Msun$ the number of stripped nuclei predicted may be too large, although the scatter in the simulations is large at the high-mass end given the small number of objects and stochastic nature of galaxy mergers.

The good agreement between the predicted number of stripped nuclei and the excess of GCs$+$UCDs above the GCLF implies that the observed number of GCs$+$UCDs can be described by a combination of the GCLF and stripped nuclear clusters.
This suggests the most massive `genuine GC'\footnote{I.e. formed through the same process as the majority of GCs and not as the nuclear star cluster of a galaxy.} has a mass of $\sim 10^{7.3} \Msun$.
This is in good agreement with the most massive GC suggested from the NGC1399 GC luminosity function \citep{Hilker:2009} and that suggested using a physically-motivated Toomre argument \citep[which is found to describe very well the mass of young massive cluster populations,][]{Adamo:2015}: Based on the most massive giant molecular clouds observed at high-redshift \citep[Toomre mass $\sim$$10^9 \Msun$ at $z>2$,][]{Genzel:2011}, \citet{Kruijssen:2014} derived a maximum GC mass at formation of $10^{7.5} \Msun$ in high-redshift environments (in particular for possible progenitors of massive galaxies like NGC1399). After taking into account mass-loss due to stellar evolution such an object would have a present day mass of $\sim 10^{7.3} \Msun$, remarkably consistent with the most massive genuine GC suggested from our analysis.

Given the low numbers involved, distinguishing between a scenario where all UCDs are genuine GCs or one where UCDs are a combination of stripped nuclei and genuine GCs is nearly impossible by numbers alone. However, a combination of the Fornax GC luminosity function and the predicted mass function for stripped nuclei shows good agreement with the observed mass function of GCs$+$UCDs in the Fornax cluster. 
For masses larger than the completeness limits there is some evidence the GC$+$UCD mass function is better fit by a combination of the GCLF and stripped nucleus mass function (6.4 and 4.9 per cent agreement within 83 and 300 kpc, respectively, according to a KS test, averaged over all SSP models) than the GCLF alone (1.6 and 1.7 per cent agreement within 83 and 300 kpc, respectively, according to a KS test, averaged over all SSP models).

\begin{figure}
  \centering
  \includegraphics[width=84mm]{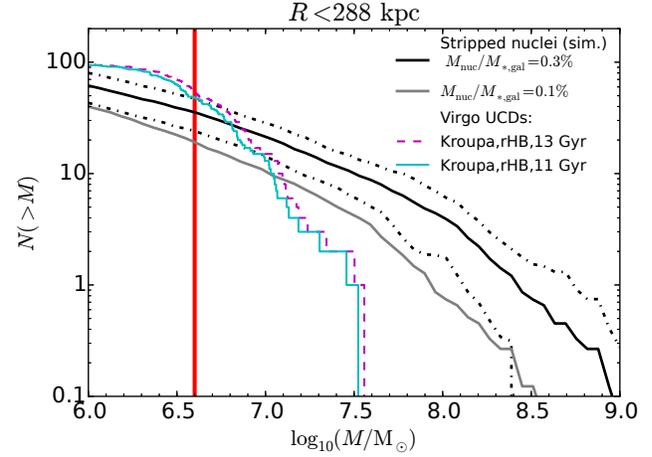}
  \caption{Cumulative mass function of simulated stripped nuclei in Virgo-sized galaxy clusters within a projected distance of 288 kpc from the central galaxy (dashed line with the standard deviation between clusters shown by dash-dotted lines) compared with UCDs ($R_e \gtrsim 10$ pc) in the Virgo cluster (thick solid line). The dashed magenta and solid cyan lines show UCDs masses using the \citet{Maraston:2005} SSP model with a Kroupa IMF for ages of 13 and 11 Gyr, respectively. The vertical red line shows the estimated completeness of the UCDs.}
  \label{plt:virgoCMF}
\end{figure}

In Fig. \ref{plt:virgoCMF} we show the cumulative mass functions of Virgo cluster UCDs and simulated stripped nuclei in Virgo-sized galaxy clusters within 288 kpc of the central galaxy (M87 for the Virgo cluster). It must be kept in mind that here UCDs are defined as having sizes $R_e \gtrsim 10$ pc while for the stripped nuclei a size cut is not possible with our model. Above masses of $\sim 10^{7.3} \Msun$ all observed GCs and UCDs have sizes larger than $\sim 10$ pc \citep[e.g.][]{MisgeldHilker:2011, Norris:2014} and therefore reasonable comparisons can be made above this mass.
For Virgo-sized clusters there are significantly more high-mass ($M > 10^{7.5} \Msun$) stripped nuclei predicted than the observed number of UCDs (although in absolute terms it is only a difference of 4-5 objects). 
Above masses of $\sim 10^{6.6} \Msun$ the total predicted number of stripped nuclei is close to the number of observed UCDs and therefore in principle stripped nuclei can account for all observed objects. However at lower masses there are more observed objects than predicted by a factor $\sim 2$, with the difference likely being even larger given the incompleteness.
The slope of the mass function predicted for stripped nuclei also differs from that of the UCDs: Over the mass range $10^{6.6}$-$10^{7.9} \Msun$ stripped nuclei have a power law slope $\alpha_\mathrm{SN} = -0.62$ while Virgo UCDs have a slope $\alpha_\mathrm{UCDs} = -1.4$.
This may be an indication that the galaxy disruption criterion is too efficient for high-mass galaxies, our assumption of a constant nucleus-to-galaxy mass ratio $M_\mathrm{nuc}/M_\mathrm{*,gal}$ with redshift is inaccurate, $M_\mathrm{nuc}/M_\mathrm{*,gal}$ of the most massive nucleated galaxies is decreased \citep[some hints of this is shown for galaxies with $M_\mathrm{*,gal} \sim 10^{11} \Msun$ in fig. 6(c) of][]{Georgiev:2016} or the nucleation fraction of galaxies with $M_\mathrm{*,gal} \sim 10^{11} \Msun$ is less than the 80 per cent we assumed. 
As we do not predict sizes for the stripped nuclei it is also possible that the most massive objects would have sizes too large to be considered UCDs (i.e. $R_e > 100$ pc) and therefore should not be included in the comparison.

\subsection{Kinematics}

\begin{figure*}
  \centering
  \includegraphics[width=0.49\textwidth]{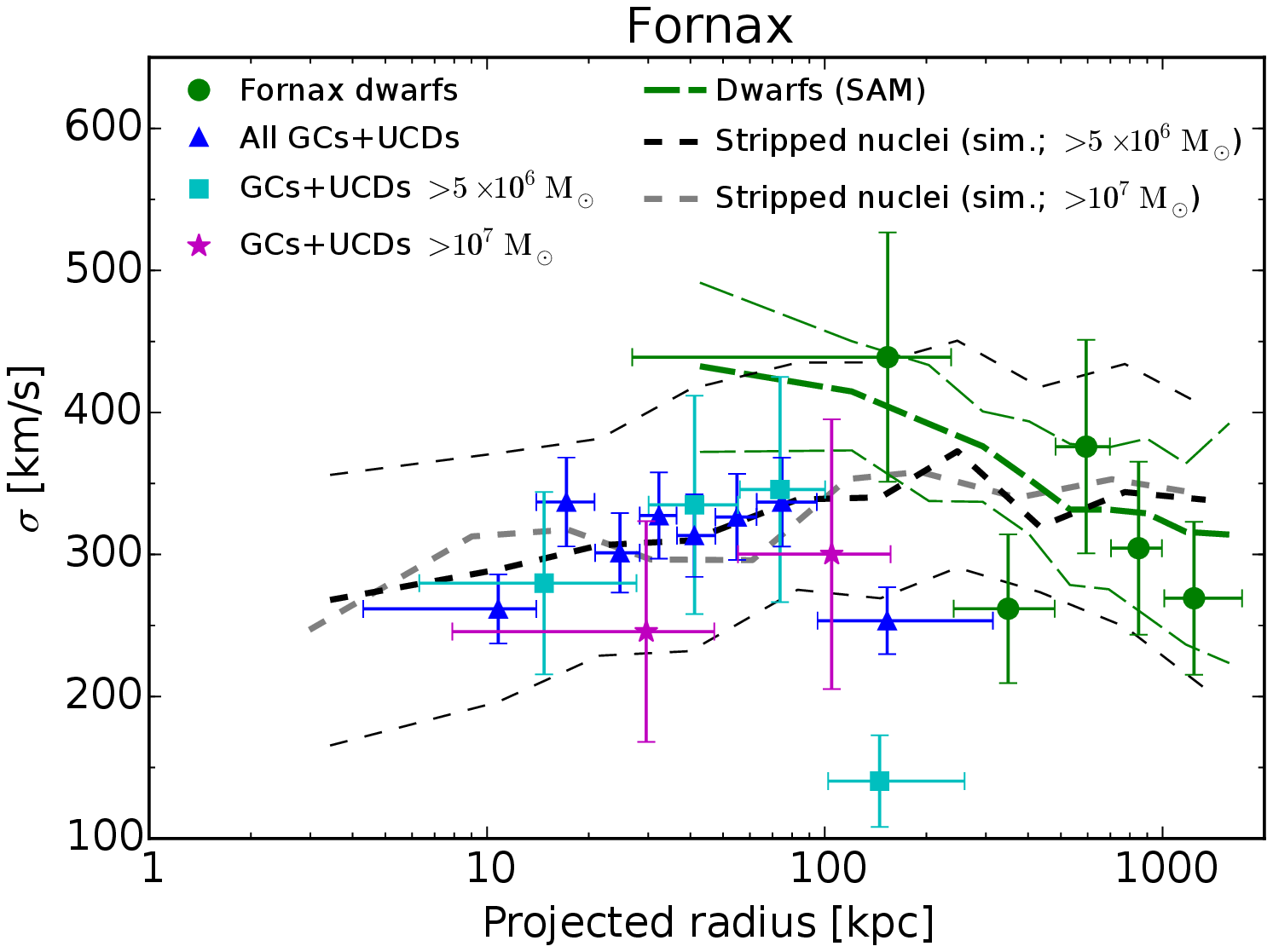}
  \includegraphics[width=0.49\textwidth]{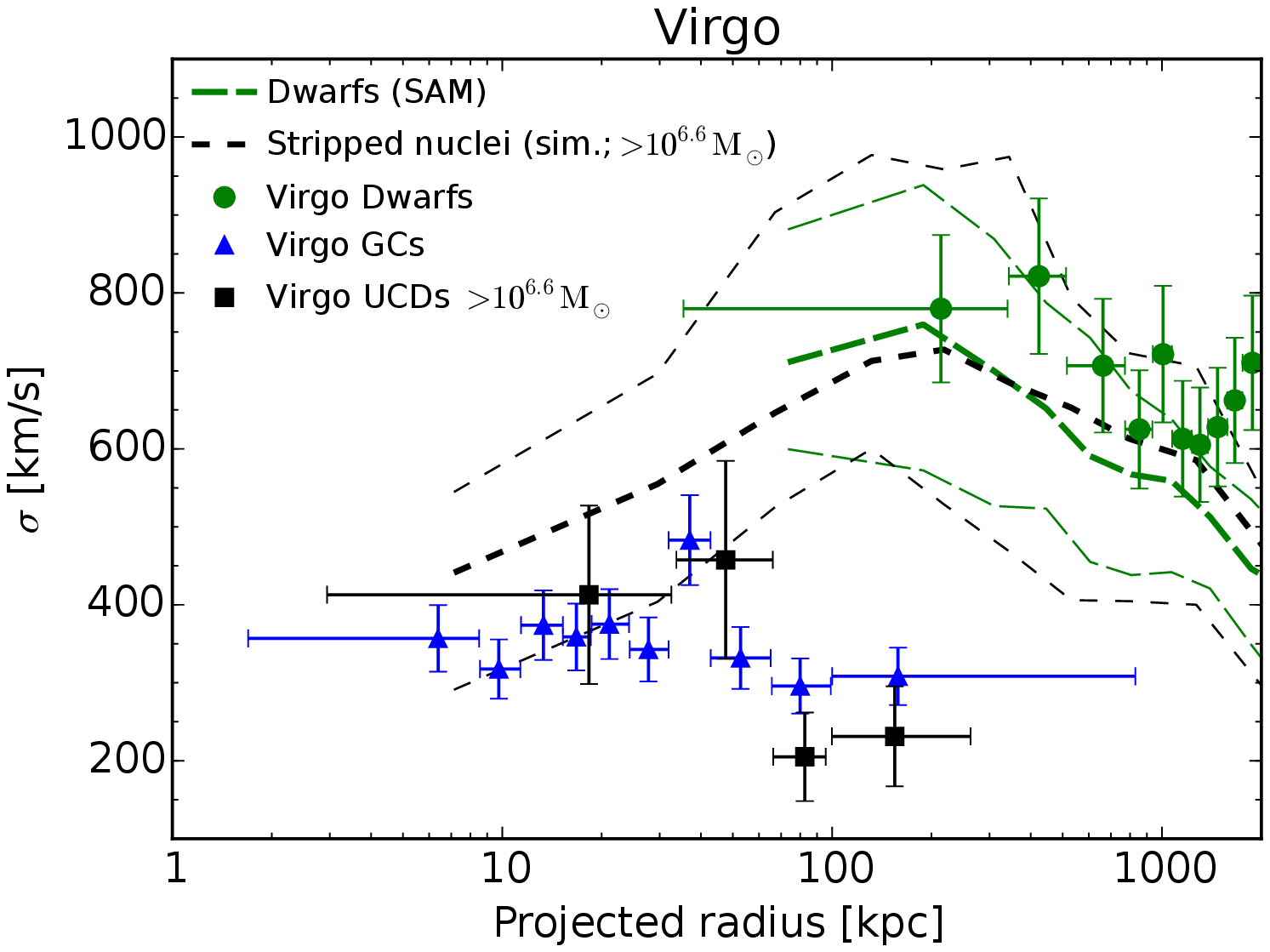}
  \caption{Velocity dispersions $\sigma$ of simulated stripped nuclei (dashed lines) and dwarf galaxies (long dashed lines) compared with GC and UCD (triangles, squares and stars) and dwarf galaxy (circles) populations in the Fornax (left panel) and Virgo clusters (right panel) as a function of projected radius. The lines for the simulation data are averages of all clusters over three sight-lines (along the $x$-, $y$- and $z$-axes of the simulation) with the standard deviation between individual clusters given by the thin lines (long and short dashes for dwarfs and stripped nuclei, respectively). All observational data is split into bins of equal numbers, where the point for radius is given by the mean of all objects within a bin and the extent of all objects within a bin is shown by the horizontal error bars. For the Fornax cluster GCs$+$UCDs we show $\sigma$ for the whole sample (blue triangles), objects with masses $M > 5\times 10^6 \Msun$ (cyan squares; approximately complete within 83 kpc) and objects with masses $M > 10^7 \Msun$ (magenta stars; approximately complete within 300 kpc). We make the same mass cuts for stripped nuclei as the observed GCs$+$UCDs (with the black and grey dashed lines for the Fornax cluster showing mass limits of $M>5\times10^6 \Msun$ and $M>10^7 \Msun$, respectively) but only show the standard deviation of the $M>5 \times 10^6 \Msun$ limit for clarity. For the Virgo cluster UCDs (defined as having $R_e \gtrsim 10$ pc) and stripped nuclei we only use objects with masses larger than $10^{6.6} \Msun$ where the observations are approximately complete within 288 kpc. Virgo cluster GCs are from \citet[excluding any objects defined as UCDs]{Strader:2011}. For the observed populations the standard deviations are determined by $\sqrt N$ statistics\protect\footnotemark[8].}
  \label{plt:velDisp}
\end{figure*}

In this section we compare the kinematic predictions for our model stripped nuclei and dwarf galaxies from the \citetalias{Guo:2011} SAM with observations in the Fornax and Virgo clusters.
Throughout this section, to determine $M/L$ ratios for observed GCs and UCDs we use the \citetalias{Maraston:2005} SSP (rHB) with a Kroupa IMF and 13 Gyr ages as the difference between the 11 and 13 Gyr models is small and the model with a Kroupa IMF is intermediate between the Chabrier and Salpeter IMF SSPs.
We caution here that since the velocities of the stripped nuclei are determined by a single particle in the simulations the velocities could be incorrect compared to that expected for a system of particles. Since we are calculating velocity dispersions and not comparing the velocities of individual nuclei we expect this will not affect our results significantly.
However the method of determining which galaxies are disrupted in the SAM may severely affect our results (e.g. by choosing galaxies with high relative velocities that may not be disrupted in reality) and therefore caution must be taken when interpreting these results.
The simulations were observed from three sight-lines (the $x$-, $y$- and $z$-axes of the simulation) and the data from all clusters were combined for the velocity dispersion measurements to improve statistics. 

\footnotetext[8]{Note that $\sqrt N$ statistics leads to underestimated errors of the velocity dispersion if the distribution even slightly deviates from normality \citep{Beers:1990} and therefore the errors are probably underestimated.}

In the left panel of Fig. \ref{plt:velDisp} we show the predicted velocity dispersions for stripped nuclei and dwarf galaxies from the simulations compared with the velocity dispersions of the Fornax GC$+$UCD and dwarf galaxy populations.
The typical deviation in average stripped nuclei velocity dispersion between individual clusters is $\sim 90$ km s$^{-1}$ and the stripped nuclei show no change of velocity dispersion with mass.
For the dwarf galaxies we take all observed galaxies with $-19.6 < M_B/\mathrm{mag} < -13.4$, corresponding to $10^8 < M_*/\Msun < 10^{10.5}$ in the simulations assuming a mass-to-light ratio $M/L_B = 3 \MLsun$, giving 25 objects per bin. 
The velocity dispersions of the SAM dwarf galaxies are in good agreement with the observed dwarf galaxies which suggests the cluster virial mass range is accurate.
For the Fornax GCs$+$UCDs we show the whole sample (117 objects per bin), objects with masses $M > 5\times 10^6 \Msun$ (approximately complete within 83 kpc; 20 objects per bin) and objects with masses $M > 10^7 \Msun$ (approximately complete within 300 kpc; 10 objects per bin). 
Due to the low number of objects per bin for the GCs$+$UCDs with masses $>5\times 10^6 \Msun$ and $>10^7 \Msun$ the actual uncertainties are probably larger than shown.
There is no significant change in the velocity dispersions of the GCs$+$UCDs with different mass cuts and is therefore consistent with the whole population originating from the same distribution. The outermost bins ($\sim 150$ kpc) for all GCs$+$UCDs and $M>5\times 10^6 \Msun$ have low velocity dispersions relative to the rest of the population. However the observations are incomplete at this radius and having a complete sample may increase the velocity dispersion measurement.

Interestingly, for the Fornax cluster the predicted velocity dispersions for stripped nuclei agree very well with the observed GCs$+$UCDs.
This may be a consequence of many GCs being accreted from dwarf galaxies by the central cluster galaxy, and thus having a similar origin and velocities to stripped nuclei \citep*[e.g.][]{Cote:1998, Hilker:1999b}.
The stripped nuclei have a velocity dispersion almost constant with radius which agrees with GCs$+$UCDs at small distances (less than 100 kpc) and dwarf galaxies at large distances (greater than 200 kpc).
As found by \citet{Gregg:2009}, UCDs and stripped nuclei have lower velocity dispersions than dwarf galaxies in the central regions of the Fornax cluster.
The difference between dwarfs and stripped nuclei implies there is a bias towards galaxies with low relative velocities (compared to the host galaxy) being disrupted at the centre of galaxy clusters.
Given the finding of similar velocity dispersions for GCs and stripped nuclei, velocity dispersions are therefore unable to distinguish between the formation mechanisms of UCDs in the Fornax cluster.

In the right panel of Fig. \ref{plt:velDisp} we show the velocity dispersions of the Virgo GC (70 objects per bin), UCD (14 objects per bin) and dwarf galaxy populations (68 objects per bin) compared with the predictions for stripped nuclei and dwarf galaxies from the simulations. 
The Virgo GCs are from the sample compiled by \citet{Strader:2011}, have $i$-band luminosities between $-7.65$ and $-12.3$ mag and are defined as having sizes $\lesssim 10$ pc. Virgo UCDs are from \citep{Zhang:2015} and are defined as having sizes $\gtrsim 10$ pc.
Again, due to the low number of objects per bin, the actual uncertainties for UCDs are probably much larger than those shown. The typical deviation in average stripped nuclei velocity dispersion between individual clusters is $\sim 130$ km s$^{-1}$.
For the observed dwarf galaxies from the EVCC we include both certain and possible members. The most distant points for the dwarfs are therefore probably significantly affected by non-cluster members that account for the rise in velocity dispersion. The velocity dispersion profile of the simulated dwarfs is on average slightly lower than the observed dwarfs but is within the cluster-to-cluster scatter of the simulations.
\citet{Zhang:2015} previously compared Virgo cluster GC and UCD velocity dispersions and found that UCDs have velocity dispersions more similar to blue GCs than red GCs and red GCs have an overall lower velocity dispersion than both blue GCs and UCDs. All three populations have dispersions lower than dwarf elliptical galaxies. Since we include both red and blue GCs the similar velocity dispersions of UCDs and GCs is due to blue GCs outnumbering red GCs by a factor of three in the sample.

For the Virgo cluster the predicted velocity dispersions of stripped nuclei match reasonably well that of the observed UCDs within 60 kpc. For larger distances the stripped nuclei have significantly elevated velocity dispersions compared to the GCs and UCDs and instead trace that of the dwarf galaxies. This could indicate that despite their sizes most Virgo UCDs are simply GCs, however more work is needed on the modelling side before strong conclusions can be made.

\begin{figure}
  \centering
  \includegraphics[width=84mm]{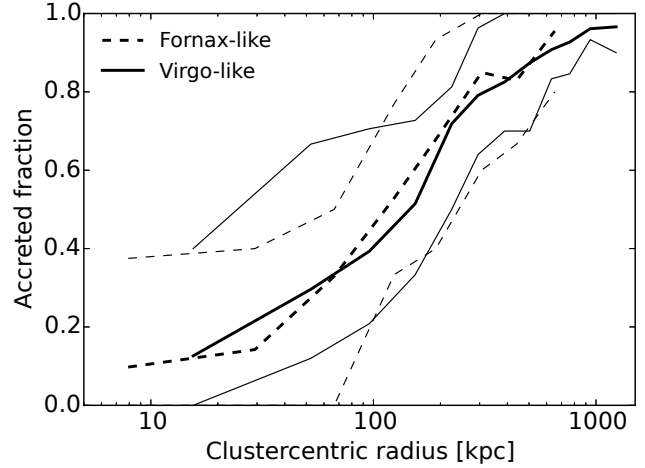}
  \caption{The accreted fraction of simulated stripped nuclei for Fornax- (thick solid line) and Virgo-like clusters (thick dashed line) as a function of clustercentric distance. The mean lines are averages of stripped nuclei in all clusters over three sight-lines (along the $x$-, $y$- and $z$-axes of the simulation) with the standard deviation between individual clusters given by the thin lines.}
  \label{plt:SNorigin}
\end{figure}

The cause of stripped nuclei velocities in Virgo-like clusters (and to a lesser extent, Fornax-like clusters) following that of the dwarf galaxies at distances larger than $\sim$100 kpc is the origin of the objects. In Fig. \ref{plt:SNorigin} we show the fraction of `accreted' stripped nuclei for Fornax- and Virgo-sized clusters. Here accreted refers to stripped nuclei that formed around galaxies other than the central galaxy of the cluster (which may or may not have then been disrupted itself). 
At distances less than 100 kpc most stripped nuclei ($>70$ per cent) form around the central galaxy of the cluster. The accreted fraction then rises sharply and above distances of 200 kpc most stripped nuclei ($>70$ per cent) are formed around satellite galaxies (or galaxies which later become satellites).
This is very similar to the distance where M87 GCs become contaminated by intra-cluster GCs \citep[$\sim 150$ kpc,][]{Zhang:2015}.

\begin{figure}
  \centering
  \includegraphics[width=84mm]{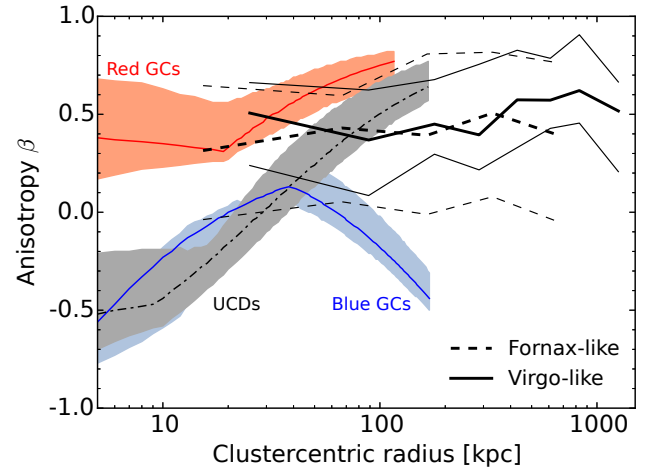}
  \caption{The orbital anisotropy of simulated stripped nuclei for Fornax- (thick solid line) and Virgo-like clusters (thick dashed line) as a function of clustercentric distance. The mean and standard deviations are as in Fig. \ref{plt:SNorigin}. We also show the anisotropy parameters determined by \citet{Zhang:2015} for Virgo cluster blue and red GCs (blue and red lines) and UCDs (black dash-dotted line), with the $1 \sigma$ confidence regions shown as the shaded areas.}
  \label{plt:anisotropy}
\end{figure}

In Fig. \ref{plt:anisotropy} we show the orbital anisotropy of stripped nuclei for Fornax- and Virgo-sized clusters. Here we include all stripped nuclei to obtain reliable statistics as anisotropy is independent of mass in our model. 
As our model does not take into account ongoing mass-loss of stripped nuclei through tidal stripping which would preferentially remove objects at small distances with radially biased orbits, at small radii ($\sim 10$ kpc) the anisotropies measured for the simulations are probably inaccurate.
We find the anisotropy of stripped nuclei is preferentially radially biased with no dependency on clustercentric distance, having a typical value of $\beta \sim 0.5$. This agrees well with the anisotropy of Virgo UCDs determined from Jeans analysis \citep{Zhang:2015} at distances of $\gtrsim 50$ kpc. At smaller radii the simulations strongly disagree with observation, where UCDs have $\beta \sim -0.5$ at 10 kpc, which can be attributed to the lack of ongoing disruption. \citeauthor{Zhang:2015} also found blue GCs have similar anisotropies to UCDs at distances smaller than 40 kpc, but have $\beta < 0$ at larger distances. This might indicate many UCDs at small distances have an origin similar to blue GCs, with the contribution of stripped nuclei increasing with distance. However our model does not include possible circularization of orbits by dynamical friction or the altering of orbits by major galaxy mergers.

\subsection{Radial distributions}

In this section we compare the predicted radial distributions for our model stripped nuclei and dwarf galaxies from the \citetalias{Guo:2011} SAM with observations in the Fornax and Virgo clusters.
As in the previous section, we use \citetalias{Maraston:2005} SSP (rHB) with a Kroupa IMF and 13 Gyr ages to determine $M/L$ ratios for the observed GCs and UCDs.

\begin{figure*}
  \centering
  \includegraphics[width=0.49\textwidth]{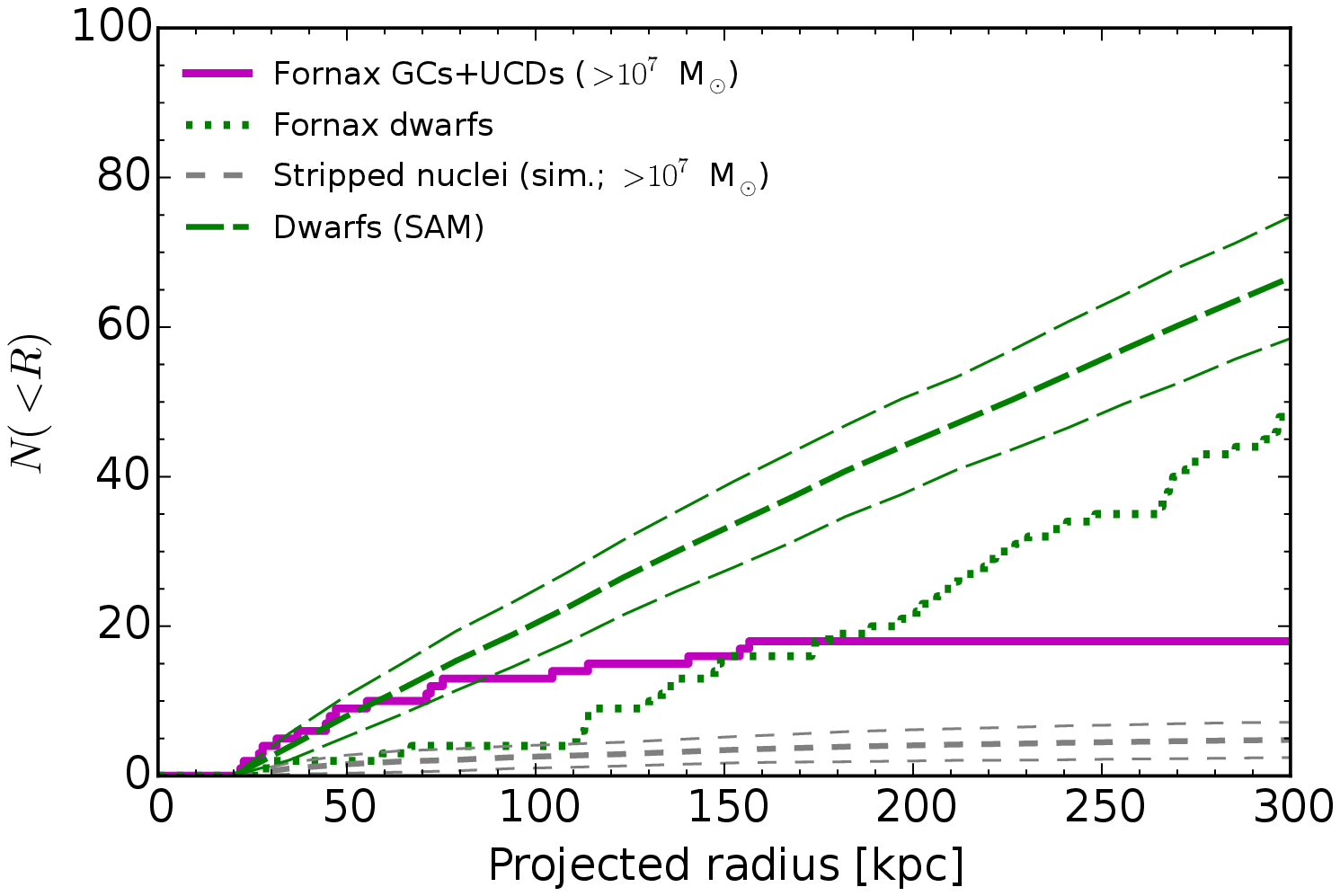}
  \includegraphics[width=0.49\textwidth]{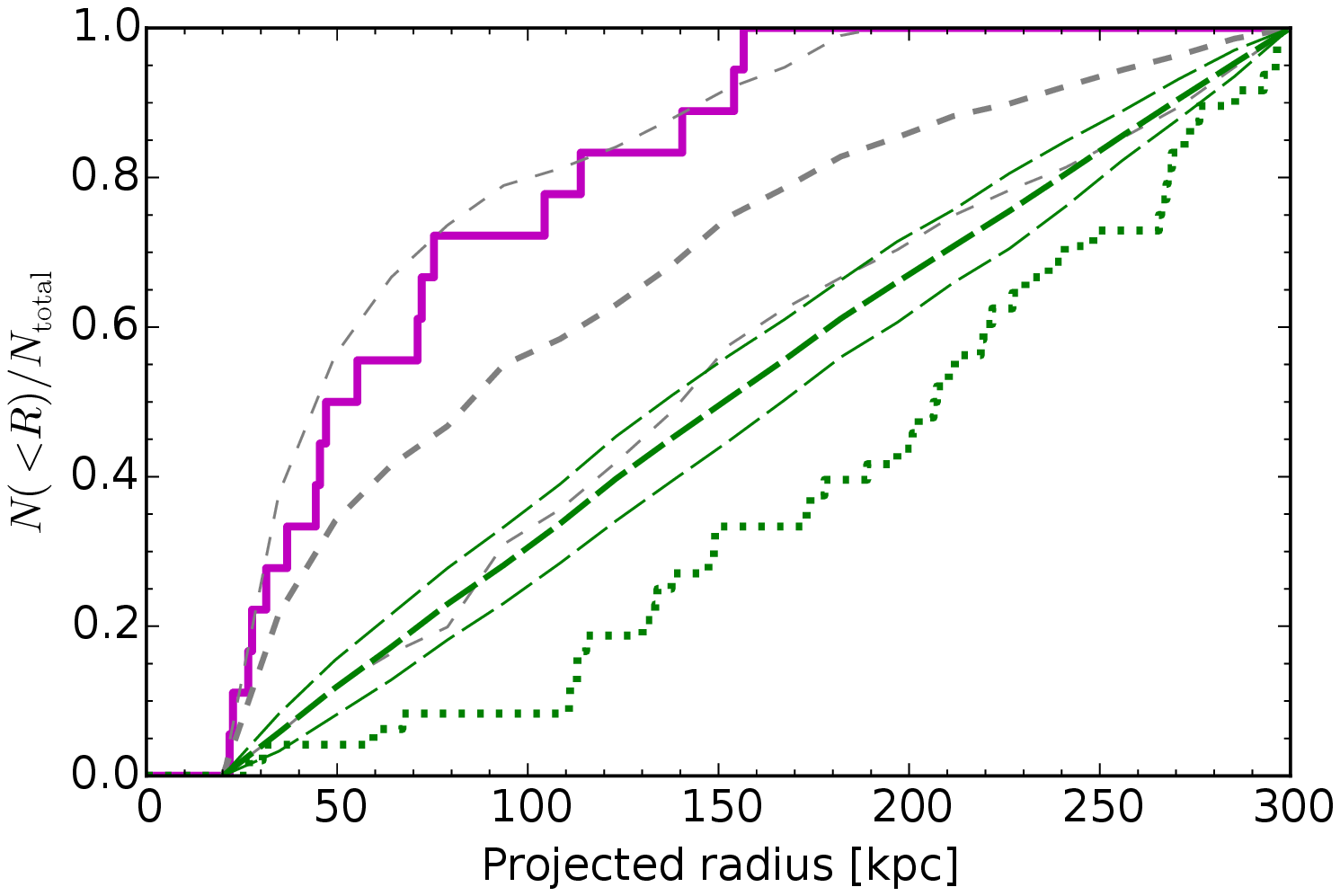}
  \caption{Absolute (left) and normalized (right) projected radial distributions of Fornax cluster GCs$+$UCDs with masses $>10^7 \Msun$ (solid magenta line) and dwarf galaxies (with luminosities $-19.6 < M_B/\mathrm{mag} < -13.4$; dotted green line) compared with the predictions of stripped nuclei with stellar masses $M_*>10^7 \Msun$ (thick grey dashed line) and dwarf galaxies (with stellar masses $10^8 < M_*/\Msun < 10^{10.5}$; thick long-dashed green line) from simulations (with standard deviation shown by the thin lines). Each simulated cluster is observed from three directions ($x$-, $y$- and $z$-axes of simulation) and the average radial distribution is used. The innermost 20 kpc is excluded since dwarf galaxy and GC$+$UCD counts are incomplete due to the high surface brightness of the central galaxy.}
  \label{plt:fornaxRadialDist}
\end{figure*}

\begin{figure}
  \centering
  \includegraphics[width=84mm]{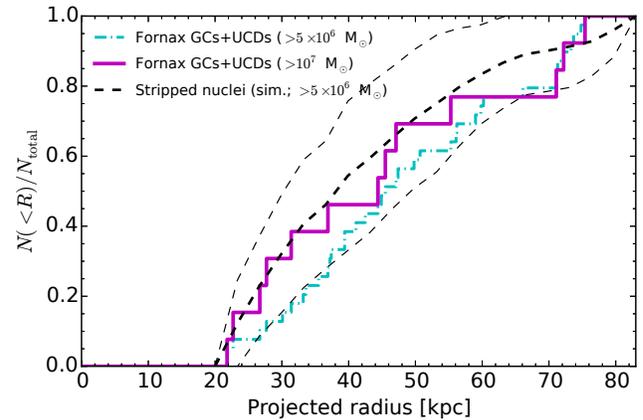}
  \caption{Normalized projected radial distribution of Fornax cluster GCs$+$UCDs with masses $>5\times 10^6 \Msun$ (dash-dotted cyan line) and $>10^7 \Msun$ (solid magenta line) and simulated stripped nuclei with masses $>5\times 10^6 \Msun$ within 83 kpc (with standard deviation shown by the thin lines).}
  \label{plt:fornaxRadialDist83}
\end{figure}

The radial distributions of GCs$+$UCDs (with masses $>10^7 \Msun$ where observations are complete) and dwarf galaxies in the Fornax cluster (solid lines) and the predictions from simulations for stripped nuclei and dwarf galaxies (dashed lines) are shown in Fig. \ref{plt:fornaxRadialDist}. 
Here we include galaxies without radial velocity measurements in the observed dwarf galaxy sample. The innermost 20 kpc is excluded since dwarf galaxy and GC$+$UCD counts are incomplete due to the high surface brightness of the central galaxy. Assuming an average mass-to-light ratio $M/L_B = 3 \MLsun$, the luminosity limits for the observed dwarf galaxies $-19.6 < M_B/\mathrm{mag} < -13.4$ approximately correspond to the mass limits for dwarf galaxies in the simulation $10^8 < M_*/\Msun < 10^{10.5}$.
The average radial distribution of the simulated stripped nuclei is more extended than that of observed GCs$+$UCDs, although, due to the low number of objects, there is large scatter in the simulations and the normalized distribution agrees with the observations at the $\sim$1-$\sigma$ level. 
In Fig. \ref{plt:fornaxRadialDist83} we show the radial distribution of Fornax GCs$+$UCDs and simulated stripped nuclei within 83 kpc, where we can be confident observations are complete. 
We don't include in the plot stripped nuclei with masses $>10^7 \Msun$ since the predictions are same as for the $>5\times 10^6 \Msun$ mass cut but with a slightly larger standard deviation.
Interestingly within this radius the radial distributions of the GCs$+$UCDs and stripped nuclei all agree very well and may suggest stripped nuclei formation is too efficient in our model beyond this distance for the most massive progenitor galaxies ($M_* \gtrsim 10^{10} \Msun$).

Both the numbers and shape of the distribution for dwarf galaxies disagrees strongly between the observations and simulations in Fig. \ref{plt:fornaxRadialDist}. This is in contrast to \citet{Weinmann:2011} who found the surface number density profile for Fornax cluster dwarf galaxies is reproduced by the \citetalias{Guo:2011} SAM. The difference is due to our fainter mass/luminosity limit for dwarf galaxies compared to \citeauthor{Weinmann:2011}.
The simulations predict 40 per cent more dwarf galaxies than observed and also show a more concentrated distribution than the observations. The over-production of dwarf galaxies in clusters in the \citetalias{Guo:2011} SAM is already well known (\citetalias{Guo:2011}; \citealt{Weinmann:2011}) and might be due to too efficient star formation in dwarf galaxies, too strong clustering on small scales or inaccurate modelling of tidal effects. 
As discussed in section 5.2 of \citetalias{Pfeffer:2014}, the too strong clustering of galaxies at small scales does not affect our predictions since the fraction of mass in a halo that was accreted in subhaloes of a given mass is relatively insensitive to the shape of the power spectrum \citep{Zentner:2003, Dooley:2014}.
Whether a change in the efficiency of star formation will affect our predictions is unclear as changes to a parameter in the SAM may be offset by other factors, although it is unlikely to affect radial distributions.
Since tidal effects are strongest in the cluster centre, more accurate modelling of tidal effects may decrease dwarf galaxy numbers and create a more extended radial distribution, thereby increasing the number of stripped nuclei.

\begin{figure*}
  \centering
  \includegraphics[width=0.49\textwidth]{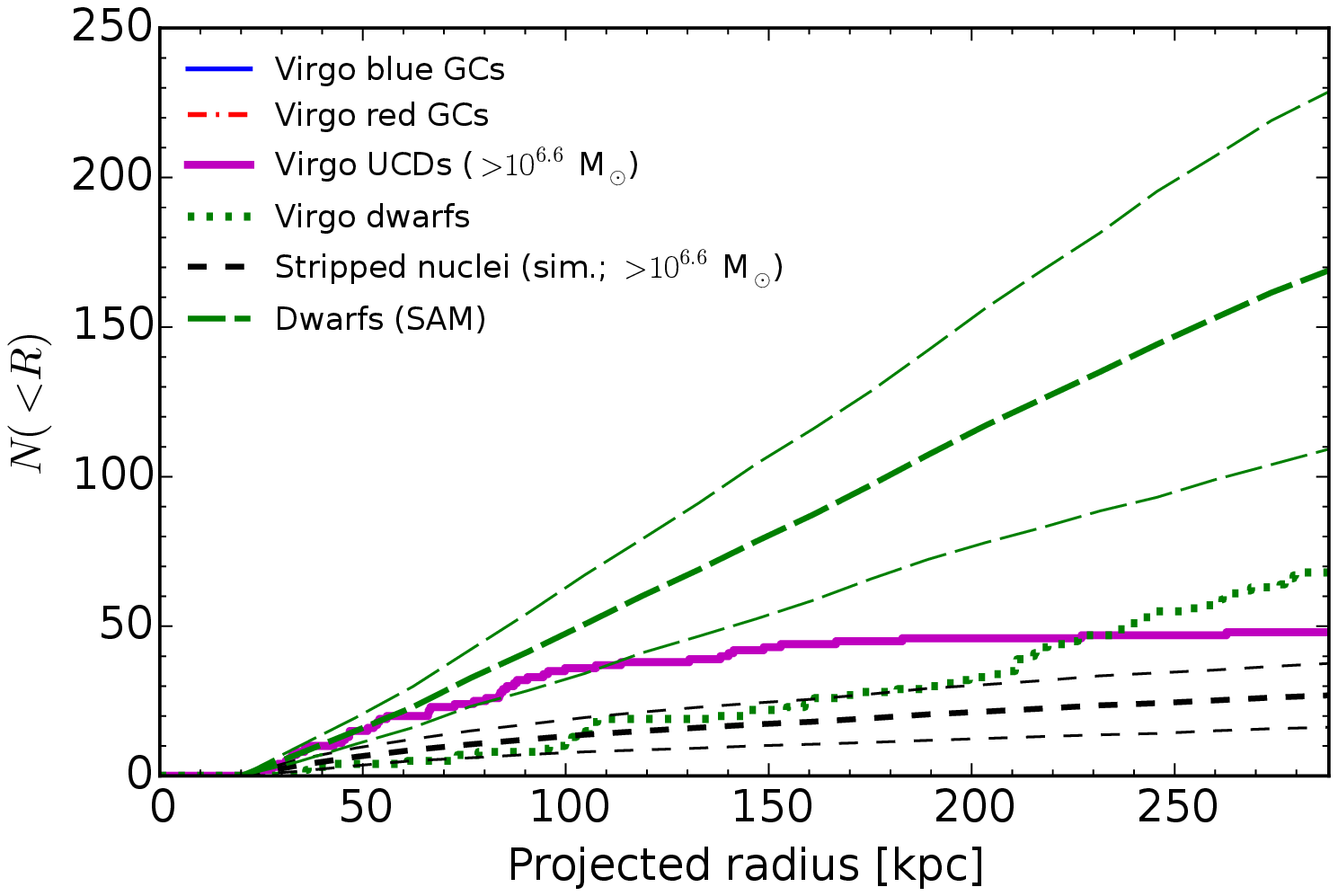}
  \includegraphics[width=0.49\textwidth]{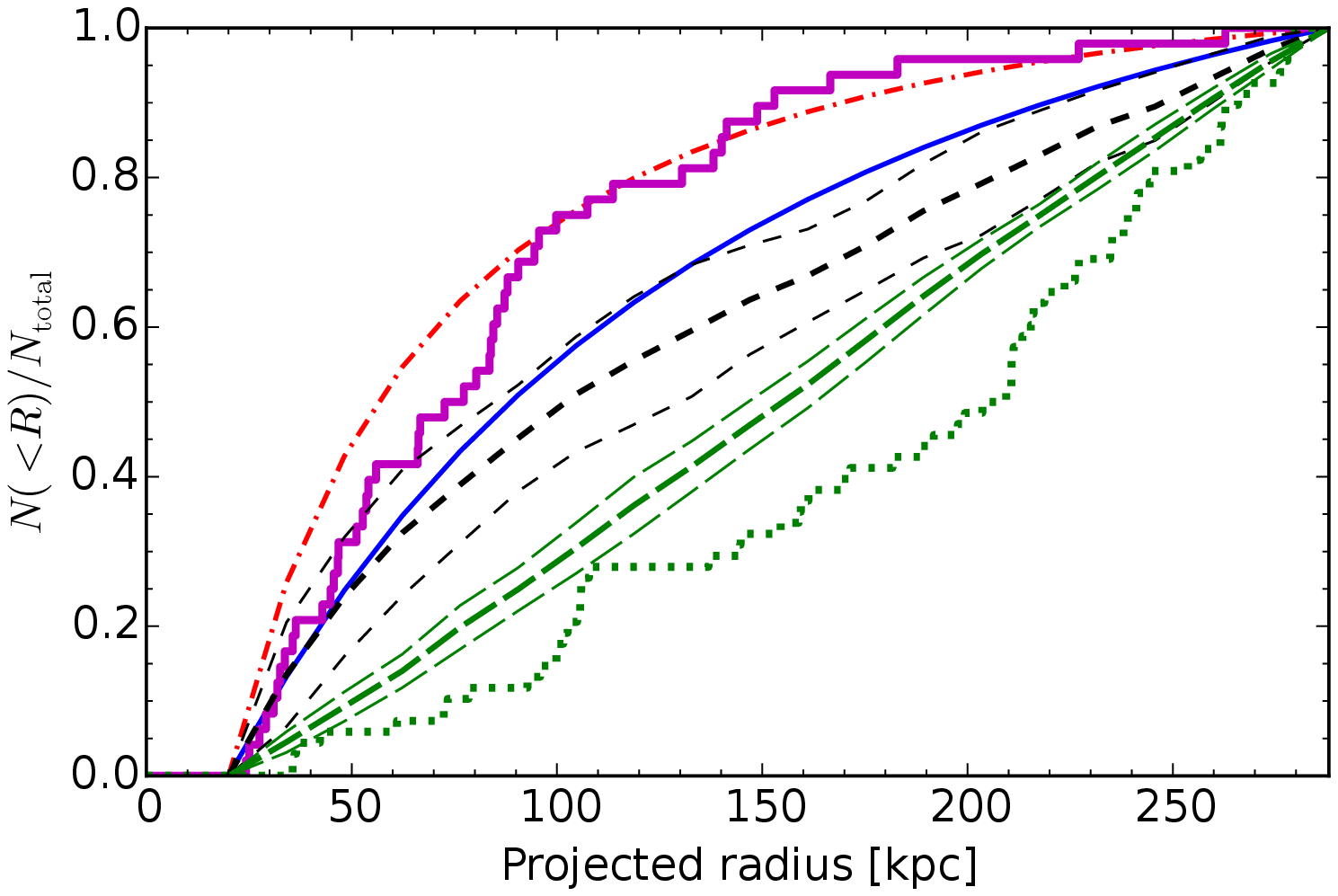}
  \caption{Absolute (left) and normalized (right) projected radial distributions of blue and red GCs (blue thin solid and red dash-dotted lines), UCDs (solid magenta line) and dwarf galaxies in the Virgo cluster (from VCC; dotted green line) within 288 kpc (within which the UCDs are approximately complete) compared with the predictions of stripped nuclei and dwarf galaxies from simulations (black dashed and green long-dashed lines with standard deviation shown by dotted lines). For the Virgo cluster UCDs and stripped nuclei we only use objects with masses larger than $10^{6.6} \Msun$ where the observations are approximately complete. Each simulated cluster is observed from three directions ($x$-, $y$- and $z$-axes of simulation) and the average radial distribution is used. The innermost 20 kpc is excluded since dwarf galaxy and GC$+$UCD counts are incomplete due to the high surface brightness of the central galaxy.}
  \label{plt:virgoRadialDist}
\end{figure*}

In Fig \ref{plt:virgoRadialDist} we show the predicted radial distributions of stripped nuclei and dwarf galaxies in Virgo-sized clusters compared with blue and red GCs, UCDs and dwarf galaxies in the Virgo cluster. The observed dwarf galaxy sample is from VCC \citep[excluding probable background galaxies]{Binggeli:1985} and includes all galaxies with masses $10^8 < M_*/\Msun < 10^{10.5}$ (assuming $M/L_B = 3 \MLsun$). As for the Fornax cluster the number of dwarf galaxies is overpredicted in the SAM and the radial distribution is too concentrated compared to the observations. The radial distribution of Virgo GCs was determined from the S\'ersic profile fits to the red and blue GCs \citep{Zhang:2015}. \citet{Zhang:2015} previously compared the surface number density profiles of red and blue GCs with UCDs. They found UCDs follow the red GCs at distances larger than 70 kpc (which can also be seen in Fig. \ref{plt:virgoRadialDist}) and have a profile shallower than both red and blue GCs at smaller distances. 

We find the stripped nuclei have a radial distribution that is significantly more extended than the UCDs and red GCs but is only slightly more extended than the blue GCs. This may suggest a common origin of stripped nuclei and blue GCs. However if more efficient disruption of galaxies at the cluster centre ($\sim$100 kpc) was implemented in the \citetalias{Guo:2011} SAM to improve dwarf galaxy numbers, stripped nuclei numbers would increase and the radial distribution would become more centrally concentrated.

\subsection{Metallicities}

\begin{figure*}
  \centering
  \includegraphics[width=0.49\textwidth]{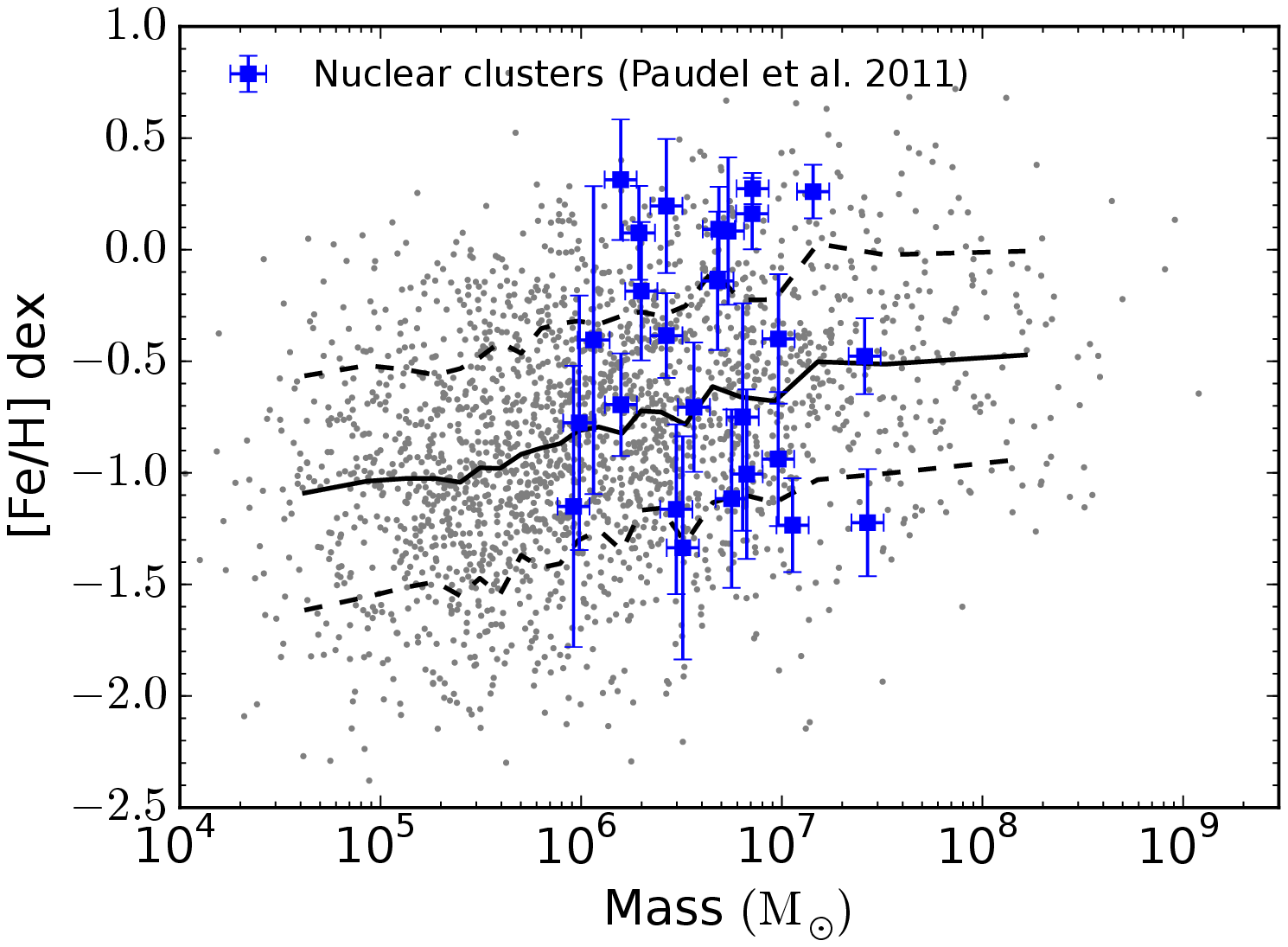}
  \includegraphics[width=0.49\textwidth]{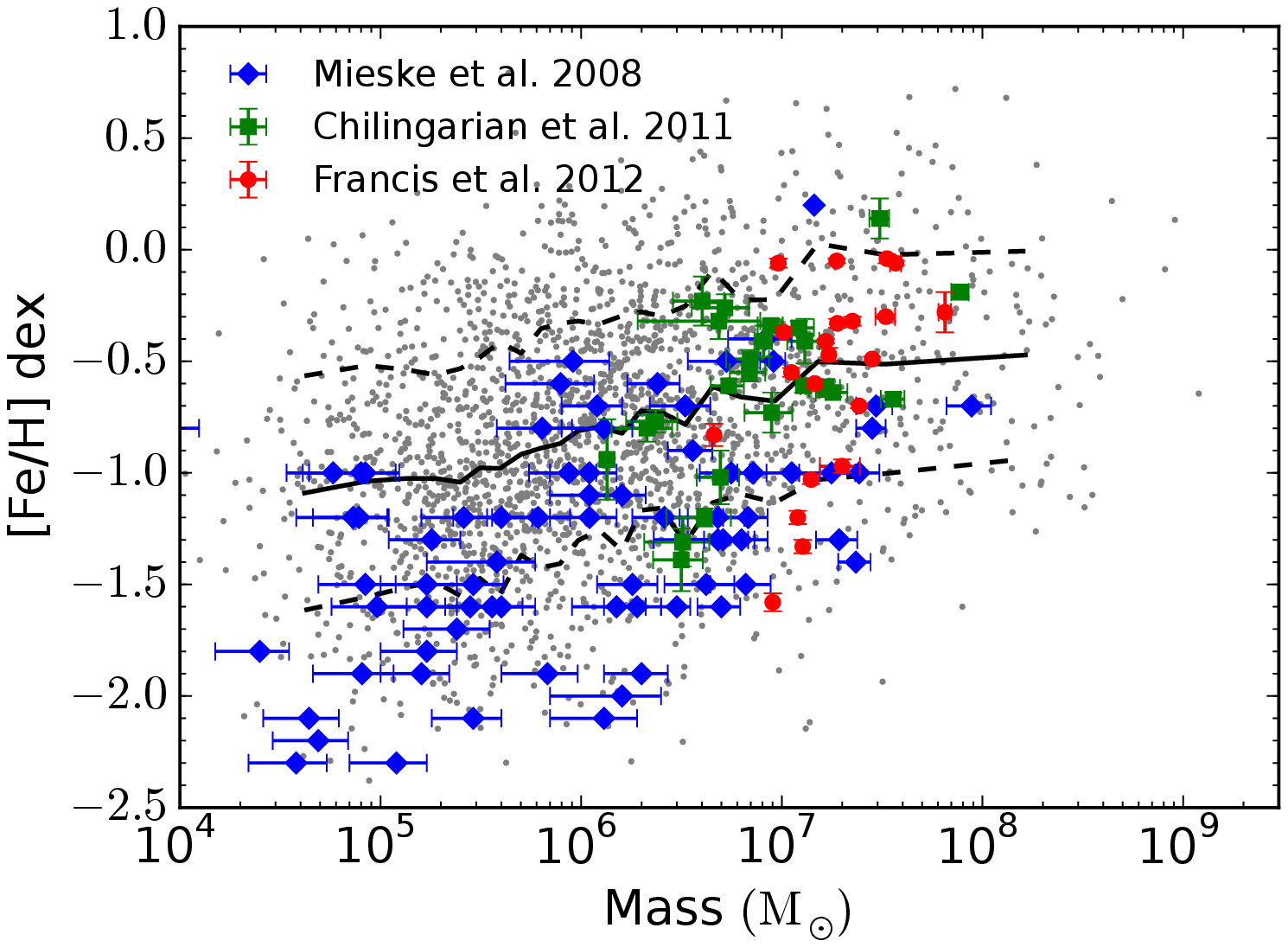}
  \caption{Predicted metallicities of stripped nuclei in Fornax- and Virgo-sized galaxy clusters (grey points) compared with nuclear clusters from \citet[left panel]{Paudel:2011} and GCs and UCDs from nearby galaxies and galaxy clusters (right panel) from \citet{Mieske:2008}, \citet{Chilingarian:2011} and \citet{Francis:2012}. The mean and confidence interval for the simulated stripped nuclei are given by the solid and dashed lines, respectively, using bin sizes of 100 objects.}
  \label{plt:metallicities}
\end{figure*}

In order to compare the predicted metallicities for stripped nuclei with the observed metallicities of UCDs we must assign metallicities to the stripped nuclei. 
We first determine $\log (\mathrm{Z}/\mathrm{Z_\odot})$ of the progenitor galaxies from the \citetalias{Guo:2011} SAM. We then convert to $\FeH$ using the relation $\ZH = \FeH + 0.94 \alphaFe$ \citep*{Thomas:2003} and assuming $\ZH = \log (\mathrm{Z}/\mathrm{Z_\odot})$.
We assume that $\alphaFe = 0.3$ which is typical of GCs, UCDs, bulges and (giant) ellipticals \citep[e.g.][]{Pritzl:2005, Francis:2012, Kuntschner:2010}.
Nuclear clusters typically have $\alphaFe \sim 0$ \citep{Paudel:2011}, although there is large scatter with some clusters having $\alphaFe \sim 0.5$. 
However nuclear clusters that can be observed are objects that have survived until the present day with potentially extended star formation histories and therefore may have different properties to nuclear clusters at high redshift. Thus it is plausible that at high redshift \citepalias[where most stripped nuclei are formed,][]{Pfeffer:2014}, nuclear clusters have elevated $\alpha$-abundance ratios due to formation in starbursts or GC infall by dynamical friction.
Finally, we assign the nucleus of each galaxy the metallicity of their host galaxy with an offset of 0.067 dex given nuclei are typically slightly more metal rich than their host galaxies and for each nucleus add a random offset drawn from a Gaussian of a width 0.46 dex to account for the typical scatter between the metallicity of nuclei and their host galaxies \citep{Paudel:2011}.
In the left panel of Fig. \ref{plt:metallicities} we show the metallicities predicted for the stripped nuclei compared with the nuclear clusters from \citet{Paudel:2011}. To convert to stellar masses, SDSS $r$-band $M/L$ ratios for the nuclear clusters were determined from their ages and metallicities with the closest fit \citetalias{Maraston:2005} models (Kroupa IMF, red horizontal branch) from their SSP grid in age and metallicity.
The metallicities we predict for the stripped nuclei are in reasonable agreement with the nuclear cluster metallicities, although within the same mass range the nuclear clusters are $0.2$ dex more metal rich on average. 

We show the predicted metallicities for stripped nuclei compared to the metallicities of GCs and UCDs (see Section \ref{sec:obsMetallicities}) in the right panel of Fig. \ref{plt:metallicities}. Where possible we use the dynamical masses measured for objects in this sample.
For masses $\gtrsim 5 \times 10^6 \Msun$ there is good agreement between the metallicities predicted for stripped nuclei and those of the observed GCs$+$UCDs, although the stripped nuclei reach larger metallicities. 
Below masses of $10^6 \Msun$ the stripped nuclei are predicted to be 0.5 dex more metal rich on average than the observed GCs. 
This is expected because many blue (i.e. metal-poor) genuine GCs mix into the sample for masses below $10^7 \Msun$ due to the mass-metallicity relation (the blue tilt) for masses above $2 \times 10^6 \Msun$ \citep[e.g.][]{Norris:2011}. 
However if nuclear clusters are mostly formed from infall of (primarily metal-poor) star clusters via dynamical friction in low-mass galaxies and star formation following gas accretion in high-mass galaxies \citep[e.g.][]{Turner:2012} the metallicity offset between nuclear clusters and their host galaxies may change with galaxy mass, causing stripped nuclei from low-mass galaxies to be more metal-poor.

\begin{figure}
  \centering
  \includegraphics[width=84mm]{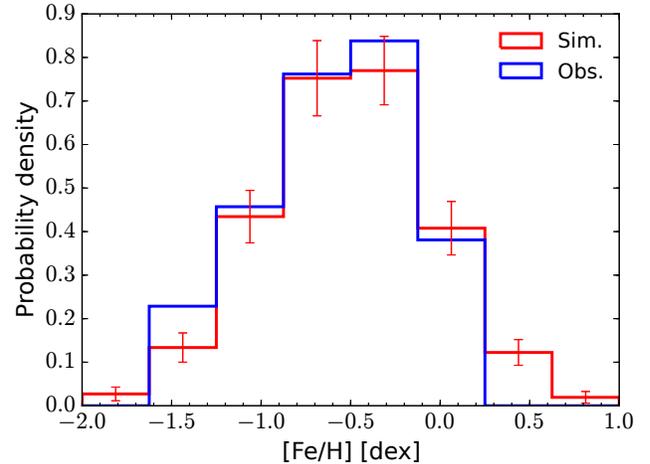}
  \caption{Normalized histogram (such that the integral under the histogram is 1) of the metallicities of observed GCs and UCDs (Obs.) and simulated stripped nuclei (Sim.) from Fig. \ref{plt:metallicities} with masses between $10^7 < M/\Msun < 10^8$. Calculations for the simulations were repeated 100 times to account for the randomly chosen metallicity offsets between galaxies and nuclear clusters with the histogram and error bars showing the mean and standard deviation for each repeat over all galaxy clusters.}
  \label{plt:metallicityHist}
\end{figure}

To compare in more detail the predictions for metallicities of stripped nuclei with those observed for GCs and UCDs, in Fig. \ref{plt:metallicityHist} we show the normalized histogram of GCs and UCDs and simulated stripped nuclei with masses between $10^7 < M/\Msun < 10^8$ for objects from Fig. \ref{plt:metallicities}. The simulation histogram shows the mean of 100 repeated metallicity assignments for the stripped nuclei to account for the randomly chosen metallicity offsets between galaxies and nuclear clusters.
We find reasonable agreement between the predictions and the observations given we have not tried to mimic the fraction of objects coming from different galaxy clusters. The four outermost bins of the simulation histogram objects show some deviation from the observed objects. However, due to the low number of stripped nuclei in each simulated galaxy cluster, the standard deviation for individual clusters is much larger than the standard deviation of the mean for all galaxy clusters and may account for this difference.
According to a KS test, the observed GCs and UCDs show a 25 per cent probability of being drawn from the same distribution as the stripped nuclei (the average in log-space of 100 repeated calculations, with a standard deviation of 0.5 dex). This shows that many of the most massive UCDs could have formed from the tidal stripping of nucleated galaxies.
However, given the trend in $\FeH$ with mass for the observed GCs and UCDs (below masses of $10^7 \Msun$ where we find stripped nuclei contribute little to the population, Section \ref{sec:massFunction}), the possibility of the most massive objects being the high-mass end of the GC population is also not ruled out by metallicity.

\subsection{Central black holes}

\begin{figure*}
  \centering
  \includegraphics[width=0.49\textwidth]{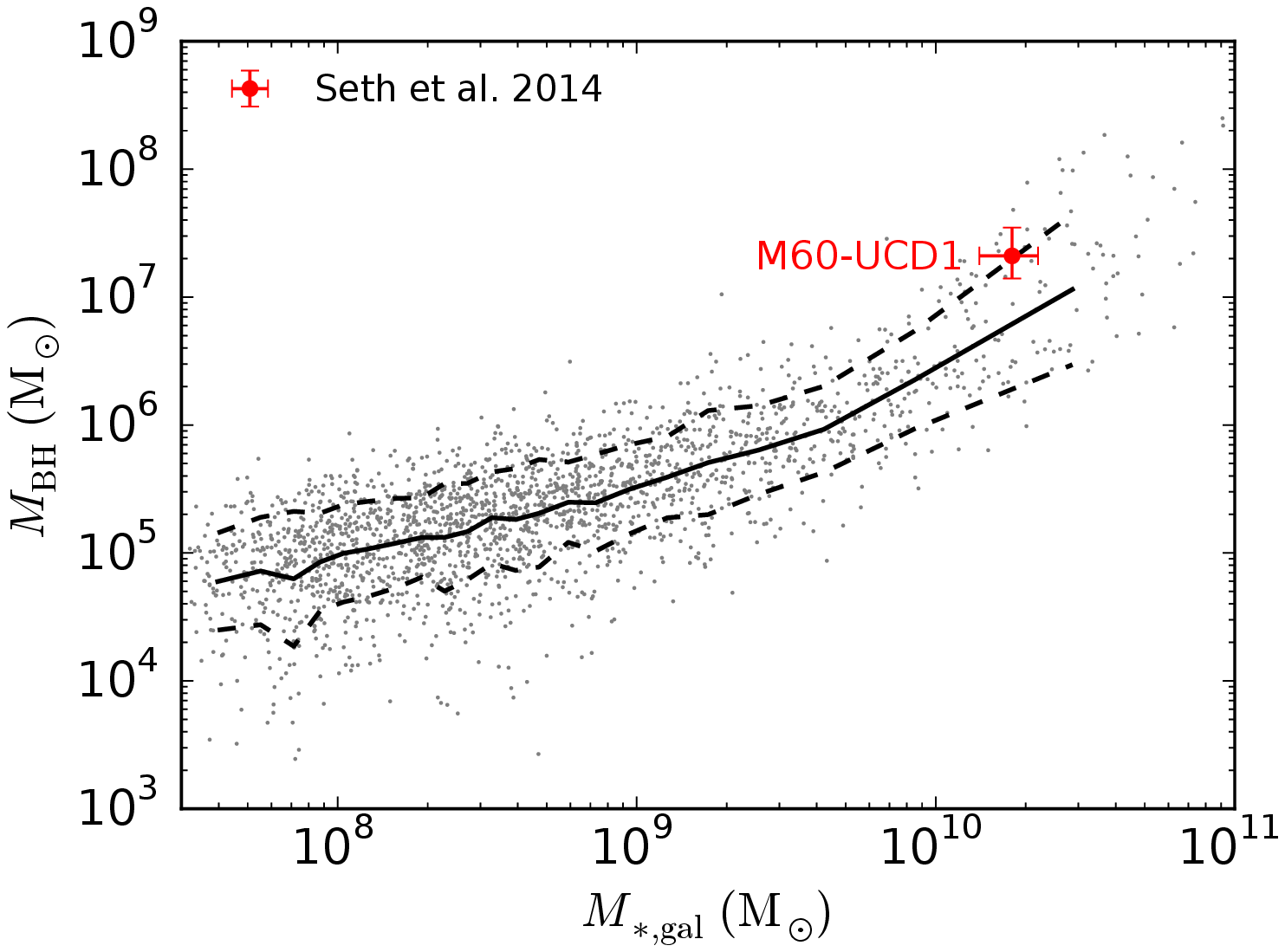}
  \includegraphics[width=0.49\textwidth]{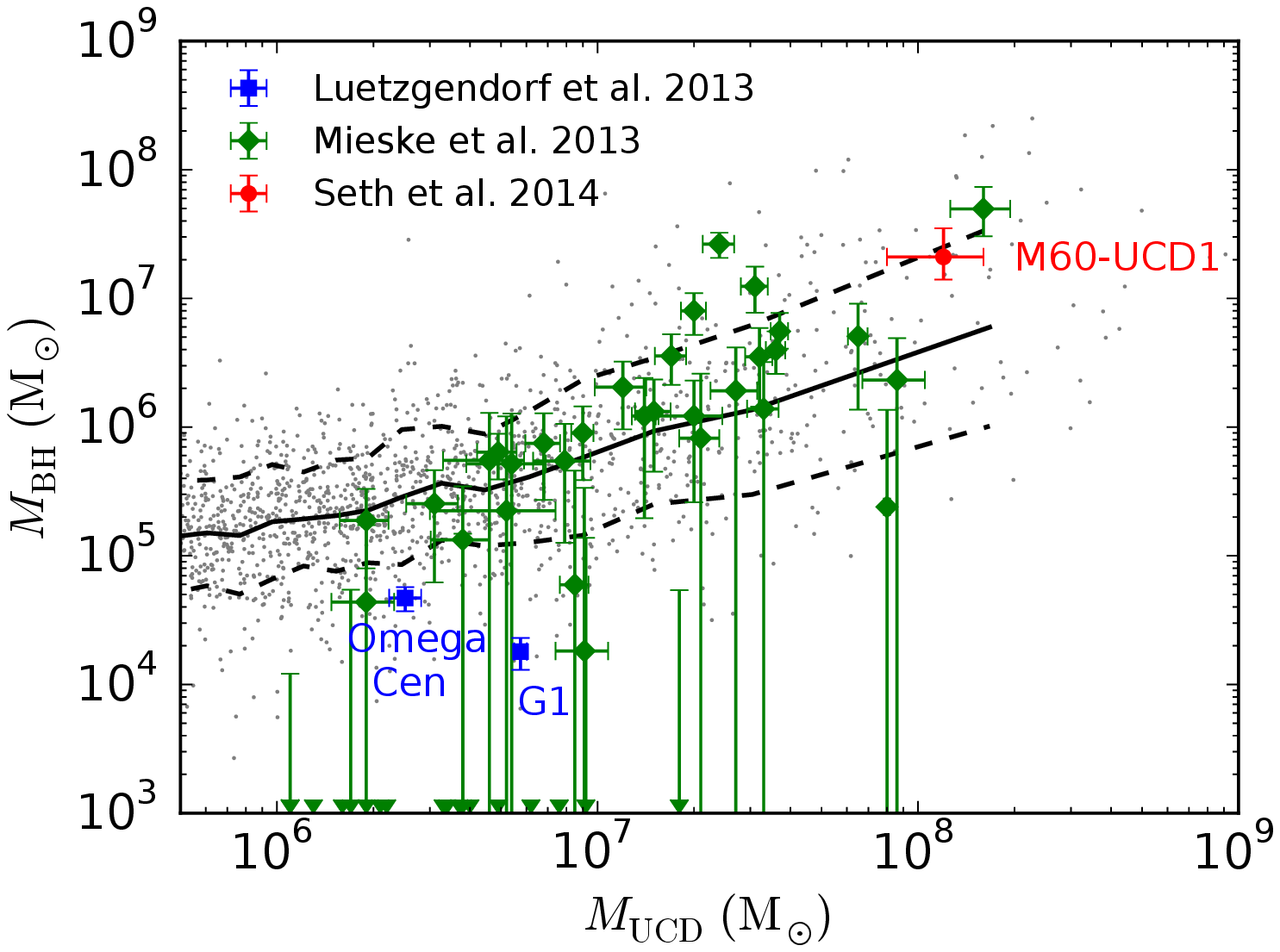}
  \caption{
Predicted masses of central black holes in stripped nuclei and their progenitor galaxy mass. The left panel plots against maximum progenitor galaxy stellar mass and the right panel compares against stripped nucleus mass. The mean and $1\sigma$ confidence interval for the simulations are given by the solid and dashed lines, respectively, using bin sizes of 100 objects. The typical $1\sigma$ confidence interval in $M_\mathrm{BH}$ is 0.5 dex. 
For comparison, in the left panel we show the black hole mass of M60-UCD1 against the estimated progenitor bulge mass \citep{Seth:2014} and in the right panel we show the black hole mass of M60-UCD1 \citep{Seth:2014}, the inferred black hole masses of UCDs assuming elevated mass-to-light ratios are due to central black holes \citep{Mieske:2013} and the limits for central black holes in the GCs $\omega$ Cen and G1 \citep{Luetzgendorf:2013}. For the observed UCDs and GCs, objects with implied black hole masses of zero are given by triangles at the bottom of the figure. 
}
  \label{plt:BHs}
\end{figure*}

UCDs are known to have dynamical mass-to-light ratios above what is expected from their stellar populations \citep{Hasegan:2005, Mieske:2008, Mieske:2013}. 
One suggestion is that these elevated mass-to-light ratios are caused by UCDs harbouring central intermediate mass black holes (IMBH) or SMBHs \citep{Mieske:2013}. The finding of the first UCD to contain a SMBH \citep[M60-UCD1,][]{Seth:2014} gives plausibility to the possibility that many UCDs may host black holes \citep[although interestingly the dynamical mass of M60-UCD1 is consistent with the stellar mass predicted from population synthesis models,][]{Strader:2013}.

In the left panel of Fig. \ref{plt:BHs} we show the predicted masses of central black holes for stripped nuclei progenitor galaxies in the \citetalias{Guo:2011} SAM.
In the right panel we show the predicted masses of central black holes in stripped nuclei, where the black hole masses are taken from the progenitor galaxies of the stripped nuclei in the SAM. 
According to the model, the fraction of stripped nuclei with a central black hole is 97 per cent. For Fornax-like clusters we predict that stripped nuclei host $13 \pm 5$ black holes more massive than $10^6 \Msun$ and $3 \pm 2$ black holes more massive than $10^7 \Msun$. For Virgo-like clusters we predict that stripped nuclei host $59 \pm 6$ black holes more massive than $10^6 \Msun$ and $14 \pm 2$ black holes more massive than $10^7 \Msun$.
In the model, black holes form during gas-rich mergers and thereafter grow either by the merger of black holes or the accretion of cold or hot gas (`quasar' and `radio' mode, respectively). 
At low galaxy masses the black hole masses are dependent on the simulation resolution and the method in which black holes are first seeded. Black hole masses in galaxies with stellar masses less than $\sim 10^9 \Msun$ (and thus stripped nuclei masses less than $\sim 10^{6.5} \Msun$) can therefore be considered very uncertain.

In Fig. \ref{plt:BHs} we also show M60-UCD1 \citep[the first UCD to have a confirmed central black hole,][]{Seth:2014}, the implied black hole masses of UCDs assuming their elevated dynamical mass-to-light ratios are due to central black holes \citep{Mieske:2013} and the limits for central black holes in the GCs $\omega$ Cen and G1 (\citealt{Luetzgendorf:2013}; both of which are thought to be stripped nuclei, e.g. \citealt{Hilker:2000}, \citealt{Meylan:2001}). 
The black hole mass of M60-UCD1 agrees well with the predictions for stripped nuclei and progenitor galaxies, falling within the $1\sigma$ confidence intervals. 
There is also remarkable agreement between the simulations and the inferred black hole masses of UCDs from \citet{Mieske:2013}. For UCDs with an implied black hole mass above zero, 74 per cent (23/31) of the data points fall within the $1\sigma$ confidence interval of the simulation predictions. For UCDs with a lower black hole mass limit above zero, 79 per cent (15/19) of the data points fall within the $1\sigma$ confidence interval of the simulation predictions. 
The GCs $\omega$ Cen and G1 both have black hole limits well below the mean predicted for their mass by the model, but are within the overall scatter of the simulated objects.


\section{DISCUSSION} \label{sec:discussion}

\subsection{The origin of UCDs}

Two main scenarios have been suggested for the formation of UCDs: they are the high-mass end of the GC mass function observed around galaxies with rich GC systems \citep{Mieske:2002, Mieske:2012} or the nuclei of tidally stripped dwarf galaxies \citep{Bekki:2001, Bekki:2003, Drinkwater:2003, Pfeffer:2013}. 
Although the internal properties (e.g. metallicities) of simulated UCDs formed by tidal stripping generally agree well with that of observed UCDs, we find tidal stripping cannot be the dominant formation mechanism of UCDs. Instead we suggest most UCDs are simply the high-mass end of the GC mass function with the contribution of UCDs formed by tidal stripping increasing towards higher masses. Above masses of $10^{7.3} \Msun$, UCDs are consistent with being entirely formed by tidal stripping. In fact for the Virgo cluster more stripped nuclei are predicted than observed UCDs by a factor of 1.5 above this mass.

It has been suggested that size and not mass differentiates between UCDs and GCs \citep[e.g.][]{Zhang:2015}. However we find that stripped nuclei cannot account for all Virgo cluster UCDs (which are defined to have sizes $\gtrsim 10$ pc). $N$-body simulations of the formation of stripped nuclei also show that for pericentre distances $\lesssim 5$ kpc (depending on the progenitor galaxy properties) the stripped nuclei formed will have sizes $<10$ pc \citep{Pfeffer:2013}.
Thus even for sizes above 10 pc GCs must make a significant contribution and size alone does not discriminate between different formation channels. 

These findings are consistent with the findings of other recent studies. \citet{Mieske:2012} calculated the fraction of GCs that contribute to the UCD population based on the specific frequencies of GCs around galaxies and found less than 50 per cent of UCDs can have formed by tidal stripping. \citet{Mieske:2013} investigated the dynamical-to-stellar mass ratios $\Psi = M_\mathrm{dyn}/M_*$ of UCDs and found evidence for a bimodal distribution, with one population having $\Psi < 1$ similar to GCs and the other having $\Psi > 1$ (see their figures 2 and 3). In the Fornax cluster less than 10 per cent (17/$\sim$200) of GCs$+$UCDs with masses $> 2 \times 10^6 \Msun$ have measured dynamical mass-to-light ratios. Of these objects, 9 have $\Psi < 1$ and are thus consistent with being genuine GCs, while 5 have $\Psi > 1$ at the $1\sigma$ level and require some additional dark mass. The remaining 3 might be either genuine GCs or UCDs formed by tidal stripping that do not host central black holes (the occupation fraction of supermassive black holes in UCD progenitor galaxies is likely not 100 per cent). In fact the only Fornax UCD with spatially resolved kinematics \citep[UCD3,][]{Frank:2011} has a dynamical mass consistent with the mass predicted from stellar population modelling \citep{Mieske:2013}.
As we find about 20 UCDs more massive than $2 \times 10^6 \Msun$ will have formed by tidal stripping in the Fornax cluster within 300 kpc (Fig. \ref{plt:fornaxCMF}), we expect most of the GC$+$UCD population without measured dynamical mass-to-light ratios should not have an elevated $\Psi$ if this scenario is correct. Therefore a combination of genuine GCs \citep[where the most massive GCs may have formed in starbursts in the early universe similar to young massive clusters, e.g.][]{Renaud:2015} and stripped nuclear clusters is currently sufficient to explain the properties of all observed UCDs in the Fornax cluster. 

If UCDs that formed by tidal stripping of nucleated galaxies host central black holes, they may make a significant contribution to the number of SMBHs in galaxy clusters. \citet{Seth:2014} found there are 45 galaxies in Fornax with stellar masses above $3 \times 10^9 \Msun$ (where SMBH occupation fraction is high) that are likely to host SMBHs. We predict for Fornax-like clusters that $14 \pm 4$ stripped nuclei with progenitor galaxy stellar masses larger than $3\times 10^9 \Msun$ may host central black holes. This would increase the number of black holes in the Fornax cluster by $\sim30$ per cent.
For Virgo-like clusters we predict $60 \pm 2$ stripped nuclei with progenitor galaxy stellar masses larger than $3\times 10^9 \Msun$ may host central black holes.

The combination of velocity dispersions ($\sigma$), anisotropy ($\beta$) and radial distributions also points to many UCDs simply being GCs. However we find conflicting results as to whether the contribution of stripped nuclei should be largest at the centre of the cluster or at increasing radii. 
\citet{Zhang:2015} found Virgo UCDs have $\beta \sim -0.5$ at $<10$ kpc, increasing to 0.5 at $>100$ kpc, and an anisotropy profile similar to blue GCs $<40$ kpc. In contrast to the UCDs, at radii larger than 40 kpc blue GCs have a tangentially biased anisotropy profile.
We predict stripped nuclei have $\beta \sim 0.5$ at all radii (Fig. \ref{plt:anisotropy}). This points to stripped nuclei contributing more with increasing radius. However effects not included in our model (circularization of orbits by dynamical friction, continuous tidal stripping removing objects on very radial orbits or major galaxy mergers altering the orbits of the innermost objects) might account for the tangentially biased orbits at the cluster centre.
On the other hand, stripped nuclei match the UCD velocity dispersions at the centre of clusters but not at distances larger than 50-100 kpc where $\sigma$ increases to match that of dwarf galaxies (Fig. \ref{plt:velDisp}), which points to stripped nuclei contributing more at the cluster centre.
Similar to the velocity dispersions, the predicted stripped nuclei radial distributions match UCDs within $\sim$100 kpc but not at larger radii. 
This may be related to the overabundance of dwarf galaxies predicted by the \citetalias{Guo:2011} SAM at cluster centres (Fig. \ref{plt:fornaxRadialDist} and \ref{plt:virgoRadialDist}) and if more realistic galaxy disruption was implemented may increase the number of stripped nuclei at the cluster centre and create a more concentrated radial distribution.
For the observed Virgo UCDs, the radially biased anisotropy and low velocity dispersion at distances of 100-200 kpc implies the number density distribution has to drop very strongly beyond these distances. This is in disagreement with the prediction for stripped nuclei where velocity dispersions increase at distances $\gtrsim100$ kpc due to the increasing contribution of stripped nuclei formed around satellite galaxies and for which there is no strong drop in number density.

The predicted velocity dispersions of stripped nuclei also has important implications for the origin of blue GCs in galaxy clusters. Blue GCs that were accreted from dwarf galaxies should have similar velocity dispersions and orbital anisotropy to stripped nuclei, i.e. $\sigma$ that follows dwarf galaxies at distances larger than 100 kpc and $\beta \sim 0.5$. The blue GCs around M87 have $\sigma \sim450$ km s$^{-1}$ at 200 kpc and a tangentially biased anisotropy \citep{Zhang:2015} which suggests a significant fraction of blue GCs could not have been accreted from dwarf galaxies around M87. Unless their orbits were significantly changed after accretion, this would imply most blue GCs were either formed in situ in the halo of M87 or possibly accreted during major mergers.

It is not clear how to reconcile the conflicting results for velocity dispersions and anisotropies with the current data. However it might imply the UCD populations are dominated by GCs and that we are not comparing objects with similar formation histories. The predicted mass function for stripped nuclei suggests the most massive UCDs have the best chance of being `genuine UCDs' (i.e. the nuclear star clusters of tidally stripped galaxies). However the number of such objects is not large enough to make statistically significant comparisons. 
Distinguishing between the formation mechanisms of individual UCDs may therefore only be possible through the presence of central black holes \citep{Seth:2014}, recent star formation \citep{Norris:2011, Norris:2015} or extended, massive tidal features of their disrupting progenitor galaxy \citep{Amorisco:2015, Mihos:2015, Voggel:2015}.

\subsection{Future work}

More work is needed on both the observation and simulation sides to further constrain the stripped nuclei scenario (and indeed the formation of UCDs generally).
In nearby galaxy clusters, complete, spectroscopically confirmed samples of GCs and UCDs to larger clustercentric radii would enable better comparisons.
Studies of different galaxy clusters have used different criteria to differentiate between GCs and UCDs; interpreting these results is difficult as different methods may obtain different results within the same cluster. Future work should investigate each galaxy cluster using consistent methods to enable comparison between clusters.
Comparisons between observed and model UCD populations should be extended beyond the central galaxy of a galaxy cluster. Programs like the NGVS \citep{Ferrarese:2012, Zhang:2015} combined with further spectroscopic surveys will enable a systematic comparison over the main subclusters of a galaxy cluster.
Observational constraints on the SMBH occupation fraction in nucleated dwarf galaxies can be used to determine what fraction of stripped nuclei (and thus UCDs) should have elevated dynamical mass-to-light ratios due to the presence of central black holes and, together with more observations of UCD dynamical mass-to-light ratios, determine if central black holes alone are sufficient to explain UCDs with elevated mass-to-light ratios.

On the simulation side a number of further refinements of our model are needed. Too many high-mass stripped nuclei are predicted for the Virgo cluster (this may also be the case for the Fornax cluster, however the number of objects is very small).
A number of possible explanations exist for this difference: 
\begin{enumerate}
\item The galaxy disruption criterion in the SAM is unrealistic. According to the SAM, galaxies are disrupted when the density of a satellite galaxy within its half-mass radius is less than the host halo density at pericentre. This might overestimate the disruption of the highest mass stripped nuclei progenitor galaxies ($M_* \sim 10^{10.5}$-$10^{11} \Msun$) since such objects would need to lose more than 99 per cent of their mass to become UCDs.
\item Related to the previous point, the high-mass stripped nuclei might not be observed as UCDs but instead as compact ellipticals. If disruption of the progenitor galaxy is incomplete (which is especially important for the highest mass progenitor galaxies that may require extremely small pericentre passages to completely remove the galaxy) the resulting object may have a size larger than 100 pc and should be considered a compact elliptical rather than a UCD.
\item Our assumption that nuclei have a constant fraction of galaxy mass at all redshifts may not be realistic.
\end{enumerate}
Future stripped nucleus formation models should include more realistic models for galaxy disruption, circularization of orbits by dynamical friction, continuous tidal stripping of the stripped nuclei and the effect of major galaxy mergers on stripped nuclei orbits.
As the build up of nuclear clusters with redshift is not possible to observe, models of nucleus formation are needed to include physically motivated nuclear cluster formation. 
Size predictions are also needed to allow for better comparisons, in particular with Virgo cluster UCDs, and to determine if the most massive stripped nuclei predicted should be included in comparisons against UCDs.
These changes may affect the predictions for mass function, velocity dispersions and radial distributions.


\section{SUMMARY} \label{sec:summary}

In this paper we compared in detail the predictions of the semi-analytic model for stripped nucleus formation presented in \citetalias{Pfeffer:2014} with the observed properties of UCDs in the local universe. Our main findings are as follows.
\begin{enumerate}
\item The number of stripped nuclei predicted at the high-mass end of the mass function in Fornax-like clusters is consistent with the most massive objects observed. For masses larger than $10^7 \Msun$ we predict stripped nuclei account for $\sim$40 per cent of GCs$+$UCDs and for masses between $10^6$ and $10^7 \Msun$ account for $\sim$2.5 per cent of GCs$+$UCDs. For the Virgo cluster more high-mass stripped nuclei are predicted than the number of observed UCDs above masses $\sim$$10^{7.4} \Msun$. Below masses of $\sim$$10^{6.7} \Msun$ more Virgo UCDs (defined as having $R_e \gtrsim 10$ pc) are observed than the predicted number of stripped nuclei.
\item The excess number of Fornax GCs$+$UCDs above the GC luminosity function agrees very well with the predicted number of stripped nuclei. Therefore the GC$+$UCD mass function is consistent with being a combination of stripped nuclei and genuine GCs (formed through the same process as the majority of GCs). This suggests the most massive genuine GC in the Fornax cluster has a mass of $\sim$$10^{7.3} \Msun$.
\item The predicted velocity dispersions for stripped nuclei in Fornax-sized clusters agrees well with the observed GC$+$UCD population at projected distances less than 100 kpc. For the Virgo cluster the velocity dispersions of stripped nuclei agree reasonably well with GCs and UCDs at distances less than $\sim$50 kpc. At larger distances the velocity dispersions of stripped nuclei increase to that of the dwarf galaxies due to the increasing contribution of stripped nuclei accreted satellite galaxies.
For both the Fornax and Virgo clusters the velocity dispersions of dwarf galaxies in the \citetalias{Guo:2011} SAM agree well with observed dwarf galaxies.
\item Stripped nuclei are predicted to have radially biased anisotropies that are approximately constant with radius ($\beta \sim 0.5$). This agrees with Virgo UCDs at distances larger than 100 kpc but disagrees at smaller distances where UCDs have $\beta$ decreasing to $-0.5$ at distances less than 10 kpc \citep{Zhang:2015}. However, ongoing disruption is not included in our model which would cause orbits to become tangentially biased at small radii.
\item For Fornax-sized clusters the radial distribution of stripped nuclei agrees well with observed GCs$+$UCDs within 83 kpc for masses $>$$5 \times10^6 \Msun$. Within 300 kpc and for masses $>$$10^7 \Msun$ stripped nuclei are on average more extended than observed GCs$+$UCDs, but consistent due to the large scatter for the simulations. For Virgo-sized clusters the radial distribution of stripped nuclei is significantly more extended than that of UCDs.
\item For masses between $10^7$ and $10^8 \Msun$, the predicted metallicities of stripped nuclei agree well with the observed metallicities of GCs and UCDs. Below masses of $10^6\Msun$, stripped nuclei are predicted to be $\sim$0.5 dex more metal-rich on average than observed GCs and UCDs.
This may be due to a metallicity offset between nuclear clusters and host galaxies that scales with galaxy mass.
\item The predicted central black hole masses for stripped nuclei agree well with the observed black hole mass of M60-UCD1 and the black hole masses implied for UCDs assuming their elevated mass-to-light ratios are due to central black holes. However, assuming Milky Way GCs with possible central black holes formed as stripped nuclei, the predictions from the modelling are 10-100 time higher than the observed values, suggesting the modelling is inaccurate at these masses or that MW GCs with IMBH estimates are not stripped nuclei. UCD black hole masses may therefore provide another approach to test SMBH formation scenarios.
\end{enumerate}

These findings together suggest that, although not the dominant mechanism of UCD formation, a significant fraction of the highest mass UCDs ($M>10^7 \Msun$) were formed by tidal stripping of nucleated galaxies.

\section*{ACKNOWLEDGEMENTS}

We thank the anonymous referee for helpful comments and Hong-Xin Zhang for providing us with data for Virgo cluster GCs and UCDs.
HB is supported by the Discovery Project grant DP110102608.
The Millennium-II Simulation databases used in this paper and the web application providing online access to them were constructed as part of the activities of the German Astrophysical Virtual Observatory (GAVO).


\label{lastpage}

\end{document}